\documentclass{article}

\usepackage{amssymb}
\usepackage{amsmath}
\usepackage{amsthm}
\usepackage{amsfonts}
\usepackage{graphicx}
\usepackage{authblk}
\usepackage{comment}

\theoremstyle{definition}

\usepackage{url}
\urldef{\mailsa}\path|{m.baldi, m.bianchi, f.chiaraluce}@univpm.it|
\urldef{\mailsb}\path|{rosenthal, davide.schipani}@math.uzh.ch|

\def\keywords{\vspace{.5em}
{\textit{Keywords}:\,\relax%
}}

\begin{document}

\title{Enhanced public key security for\\ the McEliece cryptosystem
\thanks{The material in this paper was presented in part at the
Seventh International Workshop on Coding and Cryptography (WCC 2011), 
Paris, France, April 2011.
The Research was supported in part by the Swiss National Science
Foundation under grants No. 132256, 149716, and in part by the MIUR project ``ESCAPADE'' (grant
RBFR105NLC) under the ``FIRB - Futuro in Ricerca 2010'' funding program. 
}}
\renewcommand\footnotemark{}

\author[1]{Marco Baldi}
\author[1]{Marco Bianchi}
\author[1]{Franco Chiaraluce}
\author[2]{\\Joachim Rosenthal}
\author[2]{Davide Schipani}
\affil[1]{Universit\`a Politecnica delle Marche, Ancona, Italy\\ \mailsa}
\affil[2]{University of Zurich, Zurich, Switzerland\\ \mailsb}

\date{}

\maketitle

\begin{abstract}
This paper studies a variant of the McEliece cryptosystem able to
ensure that the code used as the public key is no longer permutation-equivalent 
to the secret code.
This increases the security level of the public key, thus opening the
way for reconsidering the adoption of classical families of codes, 
like Reed-Solomon codes, that have been longly excluded from the
McEliece cryptosystem for security reasons.
It is well known that codes of these classes are able to yield
a reduction in the key size or, equivalently, an increased 
level of security against information set decoding; so, these
are the main advantages of the proposed solution.
We also describe possible vulnerabilities and attacks related
to the considered system, and show what design choices are best
suited to avoid them.
\end{abstract}


\section{Introduction}

The McEliece cryptosystem \cite{McEliece1978} is one of the most promising 
public-key cryptosystems able to resist attacks based on quantum computers.
In fact, differently from cryptosystems exploiting integer factorization
or discrete logarithms, it relies on the hardness of decoding 
a linear block code without any visible structure \cite{Berlekamp1978}.

The original McEliece cryptosystem adopts the generator matrix of a binary Goppa code as
the private key, and exploits a dense transformation matrix and a permutation matrix to 
disguise the secret key into the public one.
It has resisted cryptanalysis for more than thirty years, since no
polynomial-time attack to the system has been devised up to now; however, the increased computing 
power and the availability of optimized attack procedures have required to update its
original parameters \cite{Bernstein2008}.

The main advantage of the McEliece cryptosystem consists in its fast encryption and 
decryption procedures, which require a significantly lower number of operations with respect
to alternative solutions (like RSA).
However, the original McEliece cryptosystem has two main disadvantages: low
encryption rate and large key size, both due to the binary Goppa codes it is based on.
When adopting Goppa codes, a first improvement is obtained through the variant proposed 
by Niederreiter \cite{Niederreiter1986}, which uses parity-check matrices instead of generator matrices.
A further reduction in the public key size can be obtained by replacing binary Goppa codes
with non-binary Goppa codes, and paying attention that polynomial enumeration is prevented \cite{Bernstein2011}.

A significant improvement would be obtained if other families
of codes could be included in the system, allowing a more efficient code design 
and a more compact representation of their matrices.
In particular, the use of Generalized Reed-Solomon (GRS) codes could yield
significant advantages. In fact, GRS codes 
are maximum distance separable codes, which ensures they achieve maximum 
error correction capability under bounded-distance decoding.
In the McEliece system, this translates into
shorter keys for the same security level, or a higher
security level for the same key size, with respect to binary Goppa codes
(having the same code rate).
In fact, Goppa codes are subfield subcodes of GRS codes
and the subcoding procedure makes them less efficient than GRS codes.
However, this also makes them secure against key recovering attacks, while
the algebraic structure of GRS codes, when exposed in the public key 
(also in permuted form), makes them insecure against attacks aimed
at recovering the secret code, like the Sidelnikov-Shestakov attack \cite{Sidelnikov1992}.

Many attempts of replacing Goppa codes with other families of codes have
exposed the system to security threats \cite{Overbeck2008}, \cite{Wieschebrink2010},
and some recent proposals based on Quasi-Cyclic and Quasi-Dyadic codes have also been 
broken \cite{Umana2010}.
Low-Density Parity-Check (LDPC) codes, in principle, could offer high
design flexibility and compact keys. However, also the use of LDPC codes
may expose the system to severe flaws \cite{Monico2000, Baldi2007, Baldi2007bis, Otmani2008}.
Nevertheless, it is still possible to exploit Quasi-Cyclic LDPC codes to design a
variant of the system that is immune to any known attack \cite{Baldi2008, Baldi2012, Baldi2013icc, Baldi2013iscc}.

The idea in \cite{Baldi2008} is to replace the permutation matrix 
used in the original McEliece cryptosystem with a denser transformation 
matrix.
The transformation matrix used in \cite{Baldi2008} is a sparse matrix and 
its density must be chosen as a trade-off between two opposite effects \cite{Baldi2013icc}: 
i) increasing the density of the public code parity-check matrix,
so that it is too difficult, for an opponent, to search for low weight codewords 
in its dual code and ii) limiting the propagation of the intentional errors, so
that they are still correctable by the legitimate receiver.
The advantage of replacing the permutation with a more general transformation is that the
code used as the public key is no longer permutation equivalent to the 
secret code.
This increases the security of the public key, as it prevents an attacker 
from exploiting the permutation equivalence when trying to recover the secret 
code structure.

We elaborate on this approach by introducing a more effective class of
transformation matrices and by generalizing their form also to the non-binary case.
The new proposal is based on the fact that there exist some classes of dense transformation 
matrices that have a limited propagation effect on the intentional 
error vectors.
The use of these matrices allows to better disguise the private key into 
the public one, with a controlled error propagation effect.
So, we propose a modified cryptosystem that
can restore the use of advantageous families of codes, like GRS codes, 
by ensuring increased public key security.

 The rest of the paper is organized as follows. In Section \ref{sec:Description}, 
we describe the proposed system, both in the McEliece and Niederreiter versions. Design issues are discussed in Section \ref{sec:Design}. In Section \ref{sec:Comparison}, a comparison
with other variants of the classic McEliece cryptosystem is developed. In Section \ref{sec:Attacks},
two kinds of attacks are considered, namely the information set decoding attack and the
attack based on a particular kind of distinguisher able to tell the public matrices from 
random ones. We will show that both these attacks can be avoided, by choosing proper values of the parameters. 
In Section \ref{sec:Complexity}, key size and complexity are computed, and then compared with other solutions. Finally, in Section \ref{sec:Conclusion}, some conclusions are drawn. 

\section{Description of the cryptosystem}
\label{sec:Description}

\begin{figure}
\begin{centering}
\includegraphics[keepaspectratio, width=120mm]{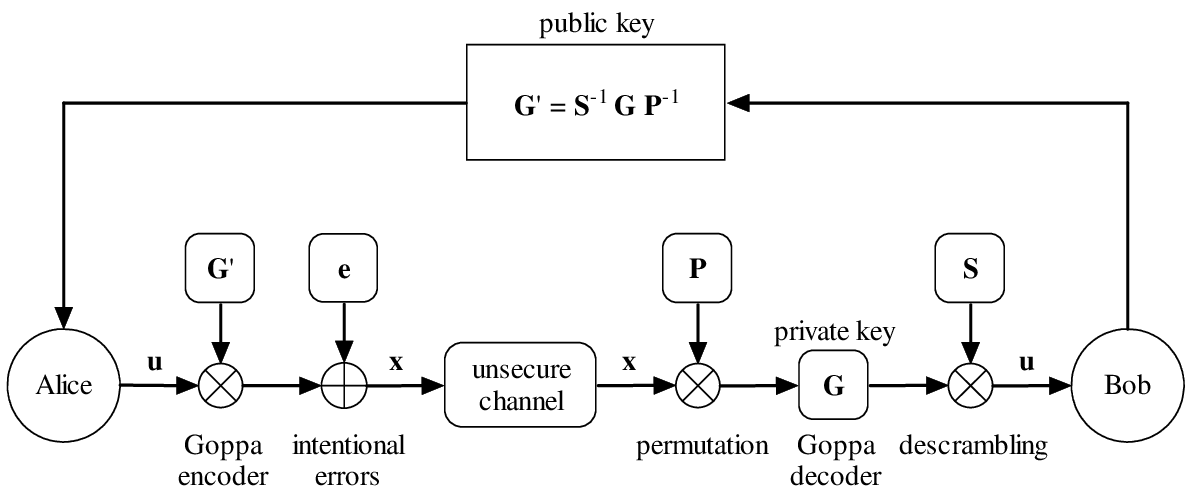}
\caption{The original McEliece cryptosystem.}
\label{fig:McEliece}
\par\end{centering}
\end{figure}

The proposed cryptosystem takes as its basis the classical McEliece cryptosystem,
whose block scheme is reported in Fig. \ref{fig:McEliece}, where $\mathbf{u}$
denotes a cleartext message and $\mathbf{x}$ its associated ciphertext. The main components 
of this system are:
\begin{itemize}
\item A private linear block code generator matrix $\mathbf{G}$
\item A public linear block code generator matrix $\mathbf{G}'$
\item A secret scrambling matrix $\mathbf{S}$
\item A secret permutation matrix $\mathbf{P}$
\item A secret intentional error vector $\mathbf{e}$
\end{itemize}
In the figure, $\mathbf{Y}^{-1}$ denotes the inverse of matrix $\mathbf{Y}$.

As for the original system, the proposed cryptosystem can be implemented
in the classical McEliece form or, alternatively, in the Niederreiter version.
In both cases, the main element that differentiates the proposed solution
from the original cryptosystem is the replacement of the permutation matrix $\mathbf{P}$
with a dense transformation matrix $\mathbf{Q}$, whose design is described next.

\subsection{Matrix $\mathbf{Q}$}
\label{subsec:Q}

The matrix $\mathbf{Q}$ is a non-singular $n \times n$ matrix having the form
\begin{equation}
\mathbf{Q} = \mathbf{R} + \mathbf{T},
\label{eq:matrixQ}
\end{equation}
where $\mathbf{R}$ is a dense $n \times n$ matrix and
$\mathbf{T}$ is a sparse $n \times n$ matrix.
The matrices $\mathbf{R}$, $\mathbf{T}$ and $\mathbf{Q}$ have elements in $\mathbb{F}_q$, with $q \geq 2$.

The matrix $\mathbf{R}$ is obtained starting from two sets, $\cal A$ and $\cal B$, each containing
$w$ matrices having size $z \times n$, $z \leq n$, defined over $\mathbb{F}_q$: 
${\cal A} = \left\{ \mathbf{a}_1, \mathbf{a}_2, \ldots, \mathbf{a}_w \right\}$,
${\cal B} = \left\{ \mathbf{b}_1, \mathbf{b}_2, \ldots, \mathbf{b}_w \right\}$.
We also define $\mathbf{a} = \sum_{i=1}^{w} \mathbf{a}_i$.
The matrices in $\cal A$ and $\cal B$ are secret and randomly chosen; then, $\mathbf{R}$
is obtained as:
\begin{equation}
\mathbf{R} = 
\left[ \begin{array}{c}
\mathbf{a}_1 \\
\mathbf{a}_2 \\
\vdots \\
\mathbf{a}_w 
\end{array} \right]^T 
\cdot \left[ \begin{array}{c}
\mathbf{b}_1 \\
\mathbf{b}_2 \\
\vdots \\
\mathbf{b}_w 
\end{array} \right],
\label{eq:matrixR}
\end{equation}
where $^T$ denotes transposition.
Starting from \eqref{eq:matrixR}, we make some simplifying assumptions, aimed
at reducing the amount of secret data that is needed to be stored.
In fact, for the instances of the proposed cryptosystem, we will focus on two distinct cases,
both with $w=2$:
i) $\mathbf{a}_1 = \mathbf{a}$, $\mathbf{a}_2 = \mathbf{0}$ and
ii) $\mathbf{b}_2 = \mathbf{1} + \mathbf{b}_1$, where $\mathbf{0}$ and $\mathbf{1}$
represent, respectively, the all-zero and the all-one $z \times n$ matrices.
In both these cases, the matrix $\mathbf{R}$ has rank $z$ and there is no need to store nor choose the matrix $\mathbf{b}_2$.
For this reason, in order to simplify the notation, we will replace $\mathbf{b}_1$ with $\mathbf{b}$
in the following.
This obviously does not prevent the applicability of the general form \eqref{eq:matrixR}
of the matrix $\mathbf{R}$.

Concerning the matrix $\mathbf{T}$, it is obtained in the form of an $n \times n$
non-singular sparse matrix having elements in $\mathbb{F}_q$ and average row and column 
weight equal to $m \ll n$, where $m$ is not necessarily an integer value.
We provide more details on its design in Section \ref{subsec:T}.

In the system we propose, the matrix $\mathbf{Q}$, having the form \eqref{eq:matrixQ},
replaces the permutation matrix $\mathbf{P}$ that is used in the original McEliece
cryptosystem and in the Niederreiter version.
All these systems exploit an intentional error vector $\mathbf{e} = \left[e_1, e_2, \ldots, e_n \right]$,
randomly generated, having a predetermined weight $t_{pub} \le t$, where $t$ is the error
correction capability of the secret code.
In the original McEliece and Niederreiter systems, $t_{pub} = t$ is used.
In the system we propose, we have instead $t_{pub} = \left\lfloor{\frac{t}{m}}\right\rfloor$.
Additionally, each error vector may be subject to further constraints, as explained below.

Let us suppose that a constraint is imposed to the vector $\mathbf{e}$ in the form:
\begin{equation}
\mathbf{a} \cdot \mathbf{e}^T = \mathbf{0}.
\label{eq:ErrorCondition}
\end{equation}
If we assume that the matrix $\mathbf{a}$ is full rank, the number of constraints
we impose, through \eqref{eq:ErrorCondition}, on the intentional error vectors is equal to $z$.
Obviously, in order to be implemented, this would require $\mathbf{a}$ to be disclosed 
as part of the public key, and this, together with
condition \eqref{eq:ErrorCondition}, may introduce a weakness in the system. 
This issue will be discussed next, together with the ways to avoid such a weakness.

For the moment, let us suppose that $\mathbf{a}$ is disclosed and that condition \eqref{eq:ErrorCondition} 
is verified.
As we will see in the following, for both the McEliece and Niederreiter versions of the cryptosystem it turns out that, 
during decryption, the matrix $\mathbf{Q}$ has a multiplicative effect on the intentional error 
vector $\mathbf{e}$.
As a result, $\mathbf{e}$ is transformed into $\mathbf{e} \cdot \mathbf{Q} =
\mathbf{e} \cdot \left( \mathbf{R} + \mathbf{T} \right)$.
If \eqref{eq:ErrorCondition} holds, 
for the two cases we focus on, the contribution due to $\mathbf{R}$ becomes:
\begin{equation}
\mathbf{e} \cdot \mathbf{R} = \left\{
\begin{array}{ll}
\mathbf{0}, 																			& \mathrm{if} \ \mathbf{a} = \mathbf{a}_1, \mathbf{a}_2 = \mathbf{0}, \\
\mathbf{e} \cdot \mathbf{a}_2^T \cdot \mathbf{1},	& \mathrm{if} \ \mathbf{b}_2 = \mathbf{1} + \mathbf{b}.
\end{array}
\right.
\label{eq:ebyR}
\end{equation}
So, in the former case, $\mathbf{e} \cdot \mathbf{Q}$ reduces to $\mathbf{e} \cdot \mathbf{T}$.
In the latter case, instead, the legitimate receiver should know the value of $\mathbf{e} \cdot \mathbf{a}_2^T$ 
to remove the contribution due to $\mathbf{e} \cdot \mathbf{R}$.
We will see in the following how this can be done.

When the result of $\mathbf{e} \cdot \mathbf{Q}$ can be reduced to $\mathbf{e} \cdot \mathbf{T}$,
the use of the matrix $\mathbf{Q}$ as in \eqref{eq:matrixQ} allows to increase the number 
of intentional errors (at most) by a factor $m$.
For $m=1$, the required error correction capability is exactly the same as in the original
McEliece and Niederreiter cryptosystems while, for $m>1$, for the same number of intentional errors, codes with higher error correction capability are required.
LDPC codes can be used for such purpose \cite{Baldi2008}, \cite{Baldi2012}.

The advantage of using the matrix $\mathbf{Q}$ is that it allows to disguise the private 
matrix of a code over $\mathbb{F}_q$ in a way that can be much stronger than by using the standard permutation matrix (as in the original McEliece system).

So, the proposed solution can help revitalizing previous attempts of using alternative families of codes in the 
McEliece system. A first challenge is to reconsider the usage of GRS codes over $\mathbb{F}_q$. 
In the following sections we will show that the attacks that have prevented their use in the past cannot
be applied to the new variant, so that it shall be considered secure against them.

\subsection{McEliece version}

In the McEliece version of the proposed system, Bob chooses his
secret key as the $k \times n$ systematic generator matrix $\mathbf{G}$ of a linear block 
code over $\mathbb{F}_q$, able to correct $t$ errors. 
He also chooses two further secret matrices: a $k \times k$ non-singular scrambling 
matrix $\mathbf{S}$ and the $n \times n$ non-singular transformation matrix $\mathbf{Q}$,
defined in \eqref{eq:matrixQ}. The public key is:
\begin{equation}
\mathbf{G}' = \mathbf{S}^{-1} \cdot \mathbf{G} \cdot{\mathbf{Q}^{-1}}.
\label{eq:Gprime}
\end{equation}
So, in general, differently from the original McEliece cryptosystem,
the public code is not permutation-equivalent to the private code.

Alice, after obtaining Bob's public key, applies the following encryption map:
\begin{equation}
\mathbf{x} = \mathbf{u} \cdot \mathbf{G}' + \mathbf{e}.
\end{equation}
After receiving $\mathbf{x}$, Bob inverts the transformation as follows:
\begin{equation}
\mathbf{x}' = \mathbf{x} \cdot \mathbf{Q} = \mathbf{u} \cdot \mathbf{S}^{-1} \cdot \mathbf{G} + \mathbf{e} \cdot \mathbf{Q},
\end{equation}
thus obtaining a codeword of the secret code affected by the error vector 
$\mathbf{e} \cdot \mathbf{Q}$.

The special form we adopt for the matrix $\mathbf{Q}$ allows Bob to reduce $\mathbf{e} \cdot \mathbf{Q}$
to $\mathbf{e} \cdot \mathbf{T}$.
Obviously, this is immediately verified when $\mathbf{e} \cdot \mathbf{R} = \mathbf{0}$ (former option in \eqref{eq:ebyR}),
while it will be shown in Sections \ref{subsec:FirstImpl} and \ref{subsec:SecondImpl}
how it can be achieved when $\mathbf{e} \cdot \mathbf{R} \ne \mathbf{0}$.

Bob is able to correct all the errors and get $\mathbf{u} \cdot \mathbf{S}^{-1}$, 
thanks to the systematic form of $\mathbf{G}$.
He can then obtain $\mathbf{u}$ through multiplication by $\mathbf{S}$.

\subsection{Niederreiter version}

The Niederreiter version of the proposed cryptosystem works as follows.
Bob chooses the secret linear block code over $\mathbb{F}_q$, able to correct $t$ errors,
by fixing its $r \times n$ parity-check matrix ($\mathbf{H}$),
and obtains his public key as
\begin{equation}
\mathbf{H'} = \mathbf{S^{-1} \cdot H \cdot Q}^T,
\label{eq:NiederreiterKey}
\end{equation}
where the scrambling matrix $\mathbf{S}$ is a non-singular $r \times r$ matrix and
the transformation matrix $\mathbf{Q}$ is defined as in \eqref{eq:matrixQ}.

Alice gets Bob's public key, maps the cleartext vector into an error vector $\mathbf{e}$ with weight $t_{pub} = \lfloor{\frac{t}{m}}\rfloor$,
and calculates the ciphertext as the syndrome $\mathbf{x}$ of $\mathbf{e}$ through $\mathbf{H'}$, according to
\begin{equation}
\mathbf{x = H' \cdot e}^T.
\label{eq:NiederreiterCiphertext}
\end{equation}

In order to decrypt $\mathbf{x}$, Bob first calculates $\mathbf{x'} = \mathbf{S} \cdot \mathbf{x} =
\mathbf{H \cdot Q}^T \cdot \mathbf{e}^T = \mathbf{H} \cdot \left(\mathbf{e} \cdot \mathbf{Q}\right)^T$.
The special form of $\mathbf{Q}$ allows Bob to reduce $\mathbf{e} \cdot \mathbf{Q}$
to $\mathbf{e} \cdot \mathbf{T}$.
Obviously, this is immediately verified when $\mathbf{e} \cdot \mathbf{R} = \mathbf{0}$,
while it will be shown in Sections \ref{subsec:FirstImpl} and \ref{subsec:SecondImpl}
how it can be achieved when $\mathbf{e} \cdot \mathbf{R} \ne \mathbf{0}$.

So, Bob gets $\mathbf{H \cdot T}^T \cdot \mathbf{e}^T$ and
he is able to obtain $\mathbf{e}_T = \mathbf{T}^T \cdot \mathbf{e}^T$, having weight $\le t$, by performing 
syndrome decoding through the private linear block code.
Then, he multiplies the result by $(\mathbf{T}^T)^{-1}$ and finally demaps $\mathbf{e}$ into its associated cleartext vector $\mathbf{u}$.

In order to reduce the public key size, the matrix $\mathbf{H'}$, defined by \eqref{eq:NiederreiterKey},
can be put in systematic form. For this purpose, let us divide $\mathbf{H'}$ into a left $r \times r$ matrix
$\mathbf{H'}_l$ and a right $r \times k$ matrix $\mathbf{H'}_r$, i.e. $\mathbf{H'} = \left[ \mathbf{H'}_l | \mathbf{H'}_r \right]$.
We can suppose, without loss of generality, that $\mathbf{H'}_l$ is full rank and obtain the
systematic form of $\mathbf{H'}$ as:
\begin{equation}
\mathbf{H''} = \left( \mathbf{H'}_l \right)^{-1} \cdot \mathbf{H'} = \left[\mathbf{I} | \left( \mathbf{H'}_l \right)^{-1} \cdot \mathbf{H'}_r \right] = \left[\mathbf{I} | \mathbf{H''}_r \right].
\label{eq:NiederreiterKeySystForm}
\end{equation}
If $\mathbf{H''}$ is used as the public key, only $\mathbf{H''}_r$ needs to be stored.
When Alice uses $\mathbf{H''}$ for encryption, she obtains a public message $\mathbf{x'' = H'' \cdot e}^T$.
Then, Bob must use $\mathbf{S}'' = \mathbf{S} \cdot \mathbf{H'}_l$, in the place of $\mathbf{S}$, in order to compute $\mathbf{x}' = \mathbf{S}'' \cdot \mathbf{x}''$.

\subsection{Design of $\mathbf{T}$}
\label{subsec:T}

As described in Section \ref{subsec:Q}, the matrix $\mathbf{T}$ is an $n \times n$
non-singular sparse matrix having elements in $\mathbb{F}_q$ and average row and column 
weight equal to $m \ll n$.

When $m$ is an integer, $\mathbf{T}$ can be simply obtained as the sum of $m$ generalized
permutation matrices, with the only constraint that their non-null elements do not overlap.
When instead $m$ is a rational value, a simple way to obtain $\mathbf{T}$ would be to design 
an almost regular matrix, having only row and column weights equal to $\left\lfloor m \right\rfloor$ or $\left\lceil m \right\rceil$,
(where $\lfloor x \rfloor$ and $\left\lceil x \right\rceil$ denote the greatest integer smaller
than or equal to $x$ and the smallest integer greater than or equal to $x$, respectively).
As an example, if $m = 1.4$, $40\%$ of the rows and columns in $\mathbf{T}$ 
could have weight equal to $2$, while the remaining $60\%$ of the rows and columns
could have weight equal to $1$.

However, if we design $\mathbf{T}$ in this way, the system must tolerate some probability 
that the weight of $\mathbf{e} \cdot \mathbf{T}$ overcomes $t$, in which case decoding fails.
By considering again $1 < m < 2$ (that will be the case of interest in the following), it is easy
to verify that this can happen when more than $\delta_t = t - t_{pub}$ errors occur at positions 
where $\mathbf{T}$ has weight-$2$ rows.
If we consider that the selected rows of $\mathbf{T}$ have disjoint supports, the failure probability
$P_f$ can be estimated as:
\begin{equation}
P_f = \sum_{i = \delta_t+1}^{t_{pub}} \frac{{t_{pub} \choose i}{n - t_{pub} \choose l - i}}{{n \choose l}},
\end{equation}
where $l$ denotes the number of weight-$2$ rows.
By taking into account the chance of non-disjoint supports of the selected rows, the actual
probability would result in a slightly smaller value.
To circumvent this problem, one of the following solutions can be adopted:
\begin{enumerate}
\item Limit to $\delta_t$ the number of columns of $\mathbf{T}$ with weight $> 1$.
If $\mathbf{T}$ has no more than $\delta_t$ columns with weight $> 1$, some or all of the weight-$2$ rows of $\mathbf{T}$  
have non-disjoint supports, and the weight of $\mathbf{e} \cdot \mathbf{T}$ is always $\le t$. 
Obviously, such columns of $\mathbf{T}$ may have weight $> \left\lceil m \right\rceil$ to reach the desired density.
This can be achieved by starting from a generalized permutation matrix and then choosing $\delta_t$ columns at random
and filling them at will.
In this process, the number of rows with weight $> 1$ should be maximized, since this is necessary to avoid attacks 
based on distinguishers, as we will see in Section \ref{Distinguisher}.
Hence, the rows of $\mathbf{T}$ should still have weight $\le \left\lceil m \right\rceil$.
\item Disclose the positions of the rows of $\mathbf{T}$ having weight $= \left\lceil m \right\rceil$, for example
by putting them in the first part of the matrix. This way, intentional error vectors having more than $\delta_t$
errors in those positions could be discarded. This has the drawback to leak some information on the intentional error vectors,
which could reduce the complexity of decoding attacks.
\item Reduce $t_{pub}$ such that $t_{pub} < \left\lfloor{\frac{t}{m}}\right\rfloor$ and $P_f$ becomes sufficiently small.
This way, however, the complexity of decoding attacks, which depends on the number of intentional error vectors, is reduced as well.
\end{enumerate}

Based on the above considerations, the first solution has to be preferred, since it allows to fix $t_{pub} = \left\lfloor{\frac{t}{m}}\right\rfloor$
and does not affect the security level.

\subsection{CCA2-secure conversions}

The McEliece and Niederreiter cryptosystem constructions described above, as well as
their original versions, offer one-way security under passive attacks, which is a
basic level of security for any public key cryptosystem.

However, in order to use these cryptosystems in practice, a stronger notion of security should
be achieved, that is, indistinguishability against adaptive chosen ciphertext attacks (IND-CCA2).
For this purpose, several conversions of the McEliece and Niederreiter cryptosystems have
been proposed in the literature, and they also apply to our case.

Classical CCA2-secure conversions work in the random oracle model \cite{Fujisaki1999, Kobara2001},
while the problem of finding efficient CCA2-secure conversions of these cryptosystem in the standard
model has been addressed more recently \cite{Dowsley2009, Dottling2012, Persichetti2012, Preetha2012, Rastaghi2013}.
The use of a CCA2-secure conversion also affects the public key size.
In fact, by adopting a classical CCA2-secure conversion in the random oracle model \cite{Kobara2001},
the public key size of the McEliece version can be reduced from $k\times n$ to $k\times r$ symbols, 
since a public generator matrix in systematic form can be used.
The same unfortunately cannot be achieved with the CCA2-secure conversions in the standard model
currently available, which are still rather unpractical and require larger public keys.

Nevertheless, using conversions which are CCA2-secure in the random oracle model has allowed 
to achieve very efficient practical implementations of the McEliece and Niederreiter systems \cite{Bernstein2013}.
On the other hand, as explained above, the Niederreiter construction yields public keys of $k\times r$ symbols,
both with and without CCA2-secure conversion.
Therefore, in Section \ref{sec:Complexity} we will consider this reduced amount of storage needed 
for the public key.

\section{System design}
\label{sec:Design}

In this section, we describe some critical aspects and possible weaknesses that must be carefully
considered in the design of the proposed system.

\subsection{Subcode vulnerability}
\label{subsec:SubcodeAttack}

When $\mathbf{a} = \mathbf{a}_1$ and $\mathbf{a}_2 = \mathbf{0}$, a possible 
vulnerability results from condition \eqref{eq:ErrorCondition}, since, in such
a case, a subcode of the public code is exposed, that is permutation-equivalent to a subcode 
of the private code.
In fact, if we refer to the Niederreiter version of the system, an attacker could 
consider the subcode generated by the following parity-check matrix:
\begin{equation} 
\label{eq:Hs}
\mathbf{H}_S = \left[ \begin{array}{c} \mathbf{H'} \\ \mathbf{a} \end{array} \right] = 
\left[ \begin{array}{c} \mathbf{S^{-1} \cdot H \cdot Q}^T \\ \mathbf{a} \end{array} \right] =
\left[ \begin{array}{c} \mathbf{S^{-1} \cdot H \cdot R}^T + \mathbf{S^{-1} \cdot H \cdot T}^T \\ \mathbf{a} \end{array} \right].
\end{equation} 

Each codeword $\mathbf{c}$ in the code defined by $\mathbf{H}_S$ must verify $\mathbf{a} \cdot \mathbf{c}^T = \mathbf{0}$.
Due to the form of $\mathbf{R}$,
this also implies $\mathbf{R}^T \cdot \mathbf{c}^T = \mathbf{0}$, so $\mathbf{H}_S$ 
defines a subcode of $\mathbf{H'}$ in which all codewords satisfy 
$\mathbf{S^{-1} \cdot H \cdot T}^T \cdot \mathbf{c}^T = \mathbf{0}$.
Hence, the effect of the dense $\mathbf{R}$ is removed and, when $\mathbf{T}$ is a permutation matrix (that is, when $m=1$),
the subcode defined by $\mathbf{H}_S$ is permutation-equivalent to a subcode of the secret code.
We notice that this is true both in the McEliece and Niederreiter versions of the cryptosystem since the parity check matrix
$\mathbf{H'}$ can always be deduced from the public generator matrix $\mathbf{G'}$.

The same vulnerability can also occur when $\mathbf{b}_2 = \mathbf{1} + \mathbf{b}$.
In fact, in this case,
\begin{equation}
\mathbf{R} = 
\left[ \begin{array}{c}
\mathbf{a}_1 \\
\mathbf{a}_2 \\
\end{array} \right]^T 
\cdot \left[ \begin{array}{c}
\mathbf{b} \\
\mathbf{1} + \mathbf{b} \\
\end{array} \right] = \mathbf{a}^T \cdot \mathbf{b} + \mathbf{a}_2^T \cdot \mathbf{1}
\label{eq:matrixR2}
\end{equation}
and
\begin{equation}
\mathbf{H} \cdot \mathbf{R}^T = \mathbf{H} \cdot \mathbf{b}^T \cdot \mathbf{a} + \mathbf{H} \cdot \mathbf{1}^T \cdot \mathbf{a}_2.
\end{equation}
So, when the private code includes the all-one codeword, that is, $\mathbf{H} \cdot \mathbf{1}^T = \mathbf{0}$,
it results $\mathbf{H} \cdot \mathbf{R}^T = \mathbf{H} \cdot \mathbf{b}^T \cdot \mathbf{a}$
and a vulnerable subcode is still defined by $\mathbf{H}_S$ as in \eqref{eq:Hs}.
For this reason, when $\mathbf{R}$ is defined as in \eqref{eq:matrixR2}, codes including the all-one
codeword cannot be used as secret codes.
For example, when a GRS code defined over $\mathbb{F}_q$ and having length $n=q-1$ is used, the all-one codeword 
is always present.
Shortened codes should be considered in order to avoid the all-one codeword.

When a GRS code is used and one of its subcodes is exposed (except for a permutation), an opponent could 
implement an attack of the type described in \cite{Wieschebrink2010}. 
It is possible to verify that, for practical choices of the system parameters, the
subcode defined by $\mathbf{H}_S$ given by \eqref{eq:Hs} is always weak against such an attack.

A similar situation occurs if LDPC codes are used as private codes, since low weight codewords could be searched in
the dual of the subcode defined by $\mathbf{H}_S$, so revealing some rows of $\mathbf{H}$ (though permuted) \cite{Baldi2007}.
Moreover, the existence of low weight codewords in the dual of a subcode of the public code
could be dangerous for the system security even when $\mathbf{H}_S$ is not available
to an attacker, since such codewords could still be searched in the dual of the public
code.
So, when dealing with LDPC codes, it is always recommended to define $\mathbf{T}$ by choosing $m>1$, in order to avoid the existence of
codewords with low weight in the dual of the public code \cite{Baldi2008}.

After having emphasized some potential weaknesses, in the following subsections we propose two implementations of the cryptosystem 
that avoid the subcode vulnerability.
We describe them with reference to the Niederreiter version of the cryptosystem,
but they can also be applied to its McEliece version.

\subsection{First implementation}
\label{subsec:FirstImpl}

A first solution to overcome the subcode vulnerability consists in maintaining
$\mathbf{a}_1 = \mathbf{a}$ and $\mathbf{a}_2 = \mathbf{0}$, 
but hiding the constraint vector $\mathbf{a}$.
Obviously, this also eliminates the need of selecting the intentional error vectors
according to condition \eqref{eq:ErrorCondition}.

We refer to the Niederreiter version of the cryptosystem and we fix, for simplicity, $z=1$,
but the same arguments can be extended easily to the general case $1 \le z \le n$.
Let us suppose that $\mathbf{a}$ 
is private and that the error vector $\mathbf{e}$
generated by Alice is such that $\mathbf{a} \cdot \mathbf{e}^T = \gamma$, with
$\gamma \in \mathbb{F}_q$. It follows that
\begin{equation}
\mathbf{R}^T \cdot \mathbf{e}^T = \gamma \mathbf{b}^T
\end{equation}
and
\begin{equation}
\mathbf{x'} = \mathbf{S} \cdot \mathbf{x} = \gamma \mathbf{H} \cdot \mathbf{b}^T + \mathbf{H \cdot T}^T \cdot \mathbf{e}^T.
\end{equation}

In this case, Bob can guess that the value of $\gamma$ is $\gamma_B$ and compute
\begin{equation}
\begin{array}{lcl}
\mathbf{x''} & = & \mathbf{x'} -  \gamma_B \mathbf{H} \cdot \mathbf{b}^T \\
& = & (\gamma - \gamma_B) \mathbf{H} \cdot \mathbf{b}^T + \mathbf{H \cdot T}^T \cdot \mathbf{e}^T.
\end{array}
\label{eq:xsecond-hiding}
\end{equation}

So, if $\gamma_B = \gamma$, Bob obtains $\mathbf{x''} = \mathbf{H \cdot T}^T \cdot \mathbf{e}^T$.
In such a case, he can recover $\mathbf{e}$ through syndrome decoding, check its weight and verify
that $\mathbf{a} \cdot \mathbf{e}^T = \gamma_B$.
Otherwise, it is $\gamma_B \neq \gamma$ and, supposing that $\mathbf{b}$ is not a valid codeword, syndrome decoding fails or returns an error vector $\mathbf{e}' \neq \mathbf{e}$. 
The latter case is extremely rare, as shown below, and can also be identified 
by Bob by checking the weight of $\mathbf{e}'$ and the value of $\mathbf{a} \cdot \mathbf{e}'^T$.
So, by iterating the procedure, that is, changing the value of $\gamma_B$, Bob is able to find the right $\gamma$.

The probability of finding a correctable syndrome $\mathbf{e}'$, 
for $\gamma_B \ne \gamma$, is very low.
In fact, since $\mathbf{b}$ is randomly chosen, when $\gamma_B \ne \gamma$ we can suppose that
the vector $(\gamma - \gamma_B) \mathbf{H} \cdot \mathbf{b}^T$ is a random $r \times 1$ vector
over $\mathbb{F}_q$. The total number of correctable syndromes is $\sum_{i=1}^{t}{n \choose i}\left(q-1\right)^i$, while the
total number of random $r \times 1$ vectors is $q^r$. So, the probability of obtaining a correctable
syndrome is:
\begin{equation}
P_e = \frac{\sum_{i=1}^{t}{n \choose i}\left(q-1\right)^i}{q^r}.
\label{eq:Pe}
\end{equation}
The value of $P_e$, for practical choices of the system parameters, is very low, as expected.
For example, by considering the set of parameters used in the original McEliece cryptosystem,
that is, $q=2$, $n=1024$, $k=524$, $t=50$, it results in $P_e \approx 10^{-65}$.

To conclude this subsection, we notice that, by using such an implementation, the complexity 
of the decryption stage is increased, on average, by a factor 
$\leq (q+1)/2$ with respect to the classical Niederreiter implementation.
In fact, the average number of decryption attempts needed by Bob is $(q+1)/2$.
However, some steps of the decryption procedure do not need to be repeated; so, an
increase in the decryption complexity by a factor $(q+1)/2$ corresponds to a pessimistic estimate.

\subsection{Second implementation}
\label{subsec:SecondImpl}

A second solution to the subcode vulnerability is to adopt the choice $\mathbf{a} = \mathbf{a}_1 + \mathbf{a}_2$,
$\mathbf{b}_2 = \mathbf{1} + \mathbf{b}$ and to preserve condition \eqref{eq:ErrorCondition}, that implies,
for Alice, the need to perform a selection of the error vectors.
In this case, according to \eqref{eq:ebyR}:
\begin{equation}
\mathbf{R}^T \cdot \mathbf{e}^T = \mathbf{1}^T \cdot \mathbf{a}_2 \cdot \mathbf{e}^T.
\label{eq:RTeT}
\end{equation}

If we fix, for simplicity, $z=1$ (but the same arguments can be extended easily to the general case $1 \le z \le n$)
and suppose to work over $\mathbb{F}_q$, the possible values of $\alpha = \mathbf{a}_2 \cdot \mathbf{e}^T$ are, obviously, $q$.
So, Bob needs to make up to $q$ guesses on the value of $\alpha$.

First, Bob computes $\mathbf{x'} = \mathbf{S} \cdot \mathbf{x} = \mathbf{H \cdot (R+T)}^T \cdot \mathbf{e}^T$.
By using \eqref{eq:RTeT}, we have:
\begin{equation}
\mathbf{x'} = \mathbf{H} \cdot \mathbf{1}^T \cdot \alpha + \mathbf{H \cdot T}^T \cdot \mathbf{e}^T.
\label{eq:xprime}
\end{equation}

We observe that, if the secret code included the all-one codeword, then $\mathbf{H} \cdot \mathbf{1}^T = \mathbf{0}$
and Bob would not need to guess the value of $\alpha$.
However, in this version of the cryptosystem, the use of codes including the all-one codeword is prevented
by the subcode vulnerability, as discussed in Section \ref{subsec:SubcodeAttack}; so, this facility cannot
be exploited.
Instead, Bob needs to make a first guess by supposing $\alpha = \alpha_B$ and to calculate
\begin{equation}
\mathbf{x''}_{\alpha_B} = \mathbf{x'} -  \mathbf{H} \cdot \mathbf{1}^T \cdot \alpha_B = \mathbf{H} \cdot \mathbf{1}^T \cdot (\alpha-\alpha_B) + \mathbf{H \cdot T}^T \cdot \mathbf{e}^T.
\label{eq:xsecond}
\end{equation}
If $\alpha_B = \alpha$, then $\mathbf{x''}_{\alpha_B} = \mathbf{H \cdot T}^T \cdot \mathbf{e}^T$; therefore, Bob can 
recover $\mathbf{e}$ through syndrome decoding, check its weight and verify that $\mathbf{a}_2 \cdot \mathbf{e}^T = \alpha_B$.
Otherwise, the application of syndrome decoding on $\mathbf{x''}_{\alpha_B}$
results in a decoding failure or in obtaining $\mathbf{e}' \ne \mathbf{e}$, for $\alpha_B \neq \alpha$.
As for the first implementation, the probability of obtaining a correctable 
syndrome $\mathbf{e}'$ is very small; so, when $\alpha_B \neq \alpha$, the decoder will end up reporting
failure in most cases.

Also in this case, the average number of decryption attempts needed
by Bob is $(q+1)/2$, and the decryption complexity increases by a factor $\le (q+1)/2$.

Concerning the subcode vulnerability, by using $\mathbf{a}_1 \ne \mathbf{a}$ and $\mathbf{a}_2 \ne \mathbf{a}$,
the matrix $\mathbf{H}_S$ as in \eqref{eq:Hs} no longer defines a subcode permutation-equivalent to a 
subcode of the secret code.
So, provided that the private code does not include the all-one codeword (for the reasons explained in Section \ref{subsec:SubcodeAttack}), the subcode vulnerability is eliminated.

Note that an attacker could try to sum two rows of $\mathbf{H}'$, hoping that one of them corresponds 
to a copy of the vector $\mathbf{a}_1$ in $\mathbf{R}$ and the other to a copy of
the vector $\mathbf{a}_2$, so that the sum of the two rows might still contain the vector $\mathbf{a}$.
If he were able to select only sums of this type, then he might be able to find a weak subcode. 
This, however, appears to be a hard task for the following reasons. 
If he adds one row with all the other rows, he would get, on average, only $r/2 = (n-k)/2$ rows containing 
the vector $\mathbf{a}$, while the other sums would contain $2 \mathbf{a}_1$ or $2 \mathbf{a}_2$;
even if he were able to select the rows corresponding to $\mathbf{a}$, the dimension of the subcode would not be 
large enough for a feasible attack \cite{Minder2007}, \cite{Wieschebrink2010}.
Furthermore, effectively obtaining $\mathbf{a}$ in the sum of two rows also depends on
how $\mathbf{H}$ is built, i.e. it may happen only if some special relations between elements 
of $\mathbf{H}$ are satisfied. Again, this has only a (small) probability to occur.
Lastly, summing pairs of rows would also imply summing pairs of rows of $\mathbf{T}^T$;
so, their (very low) weight would be doubled with a very high probability, making decoding harder.

For these reasons, it seems not easy to devise a further vulnerability for the
subcode that may allow to mount an attack against this implementation.

\subsection{Choice of $\mathbf{Q}$}
\label{subsec:ChoiceQ}

Also the choice of the matrix $\mathbf{Q}$ can involve some critical aspects.
Let us focus on the binary case ($q = 2$) and consider a particular instance of the 
first implementation, in which the matrix $\mathbf{Q}$ is obtained as
\begin{equation}
\mathbf{Q}_1 = \mathbf{R} + \mathbf{P}_1,
\label{eq:Q1}
\end{equation}
$\mathbf{P}_1$ being a permutation matrix and
\begin{equation}
\mathbf{R} = \mathbf{a}^T \cdot \mathbf{b} = \left[\begin{array}{cccc} {a_{1} } & {a_{2} } & {\cdots } & {a_{n} } \end{array}\right]^T \cdot \left[\begin{array}{cccc} {b_{1} } & {b_{2} } & {\cdots } & {b_{n} } \end{array}\right],
\label{eq:Rmatrix}
\end{equation}
where $\mathbf{a}$ and $\mathbf{b}$ are two random vectors over $\mathbb{F}_2$.

In the choice of $\mathbf{Q}_1$, it is important to avoid some special cases which could
allow an attacker to derive a code that is permutation-equivalent to the secret one, thus bringing
security back to that of the classical McEliece system.

For exploring the subject, let us suppose that the $j$-th element of $\mathbf{b}$ is zero and that $\mathbf{P}_1$
has a symbol $1$ at position $(i, j)$. In this case, the $j$-th column of $\mathbf{Q}_1$
is null, except for its element at position $i$.
Since $\mathbf{Q}_1^{-1} = \widehat{\mathbf{Q}_1}/\left|\mathbf{Q}_1\right|$, where
$\widehat{\mathbf{Q}_1}$ is the adjoint matrix and $\left|\mathbf{Q}_1\right|$ is the
determinant of $\mathbf{Q}_1$, it follows from the definition of $\widehat{\mathbf{Q}_1}$
that the $i$-th column of $\mathbf{Q}_1^{-1}$ is null, except for its element 
at position $j$.
So, the $i$-th column of $\mathbf{Q}_1^{-1}$ has the effect of a column
permutation, like in the original McEliece cryptosystem.

In order to avoid such a possible flaw, we impose that all the elements of $\mathbf{b}$ are non-zero.
If we limit to the binary case, this imposes that $\mathbf{b}$ is the all-one vector.
However, in such a case, further issues exist in the design of $\mathbf{Q}$.
For example, let us consider $\mathbf{a}$ as an all-one vector too, so that $\mathbf{R} = \mathbf{1}$.
A valid parity-check matrix for the public code is:
\begin{equation}
\mathbf{H}' = \mathbf{H \cdot Q}^T,
\end{equation}
where $\mathbf{H}$ is the parity-check matrix of the private code.
In the special case of $\mathbf{Q}_1 = \mathbf{1} + \mathbf{P}_1$, we have 
$\mathbf{H}' = \mathbf{H \cdot 1} + \mathbf{H \cdot P}_1^T$.
By assuming a regular $\mathbf{H}$ (i.e. with constant row and column weights),
two cases are possible:
\begin{itemize}
\item If the rows of $\mathbf{H}$ have even weight, $\mathbf{H \cdot 1} = \mathbf{0}$ and $\mathbf{H}' = \mathbf{H \cdot P}_1^T$.
\item If the rows of $\mathbf{H}$ have odd weight, $\mathbf{H \cdot 1} = \mathbf{1}$ and $\mathbf{H}' = \mathbf{1} + \mathbf{H \cdot P}_1^T$.
\end{itemize}
In both cases, the public code has a parity-check matrix that is simply a permuted version of that of the secret 
code (or its complementary).
This reduces the security to that of the original McEliece cryptosystem, that discloses a permuted version of the secret code.
Such a security level is not sufficient when adopting, for example, LDPC codes, since the permuted version of the secret matrix $\mathbf{H}$ can be attacked by searching for low weight codewords in the dual of the secret code.

A more general formulation of the flaw follows from the consideration that $\mathbf{Q}_1 = \mathbf{1} + \mathbf{P}_1$
has a very special inverse.
First of all, let us consider that $\mathbf{Q}_1$ is invertible only when it has even size. This is obvious 
since, for odd size, $\mathbf{Q}_1$ has even row/column weight; so, the sum of all its rows is the zero vector.
If we restrict ourselves to even size $\mathbf{Q}_1$ matrices, it is easy to show that their inverse 
has the form $\mathbf{Q}_1^{-1} = \mathbf{1} + \mathbf{P}_1^T$, due to the property of permutation matrices 
(as orthogonal matrices) to have their inverse coincident with the transpose.

So, $\mathbf{Q}_1^{-1}$ has the same form of $\mathbf{Q}_1$ and, as in the case of $\mathbf{H}$, 
disclosing $\mathbf{G}' = \mathbf{S}^{-1}\mathbf{GQ}_1^{-1}$ might imply disclosing a generator
matrix of a permuted version of the secret code or its complementary (depending on the parity of its row weight).
Therefore, the form $\mathbf{Q}_1 = \mathbf{1} + \mathbf{P}_1$ might reduce the security to that of
the permutation used in the original McEliece cryptosystem.

Based on these considerations, one could think that adopting a vector $\mathbf{a}$ different from the 
all-one vector could avoid the flaw.
However, by considering again that $\mathbf{Q}_1^{-1} = \widehat{\mathbf{Q}_1}/\left|\mathbf{Q}_1\right|$,
it is easy to verify that a weight-$1$ row in $\mathbf{Q}_1$ produces a weight-$1$ row in $\mathbf{Q}_1^{-1}$
and a weight-$(n-1)$ row in $\mathbf{Q}_1$ produces a weight-$(n-1)$ row in $\mathbf{Q}_1^{-1}$.
It follows that $\mathbf{Q}_1^{-1}$ contains couples of columns having Hamming distance $2$.
Since their sum is a weight-$2$ vector, the sum of the corresponding columns of the public matrix 
results in the sum of two columns of $\mathbf{S}^{-1}\mathbf{G}$.
Starting from this fact, an attacker could try to solve a system of linear equations with the
aim of obtaining a permutation-equivalent representation of the secret code, at least for the
existing distance-$2$ column pairs.

So, our conclusion concerning the binary case is that the choice of $\mathbf{Q}$ as in \eqref{eq:Q1}
and \eqref{eq:Rmatrix} should be avoided. 
A safer $\mathbf{Q}$ is obtained by considering $z > 1$ and $m > 1$. 
This obviously has the drawback of requiring codes with increased error correction capability.

These considerations about the structure of the matrix $\mathbf{Q}$ are useful in general for every family of
codes we would like to use in the proposed system, but a more specific characterization is needed depending
on the type of codes adopted. In fact, in order to avoid specific attacks or to choose feasible parameters
linked to the code's error correction capability, it is necessary to address further structure issues like those
we will analyze in Section \ref{subsec:Distinguisher}.   

\section{Comparison with other variants of the McEliece cryptosystem}
\label{sec:Comparison}

The main difference between the proposed cryptosystem and many other variants of
the McEliece cryptosystem consists in the way the secret generator matrix is 
disguised into the public one, that is, by using a more general transformation matrix in
the place of the permutation matrix.

Other proposals for increasing key security have been made in the past,
such as using a distortion matrix together with rank codes in the GPT cryptosystem
\cite{Gabidulin1991} and exploiting the properties of subcodes in variants of
the McEliece and the GPT cryptosystems \cite{Berger2005}.
Unfortunately, cryptanalysis has shown that such approaches exhibit security
flaws \cite{Overbeck2008}, \cite{Wieschebrink2010}.

The idea of using a rank-$1$ matrix with the structure \eqref{eq:Rmatrix}
can be found in \cite{Gabidulin1994}. However, such a matrix was added to the secret 
matrix (rather than multiplied by it) and no selection of the error vectors was 
performed, so that a completely different solution was implemented.

Instead, the idea of replacing the permutation in the McEliece cryptosystem with a more
general transformation matrix is already present in the variant of the GPT cryptosystem
adopting a column scrambler \cite{Ourivski2003}, \cite{Rashwan2010} and in cryptosystems 
based on full decoding \cite[sec. 8.3]{Kabatiansky2005}.
These proposals are shortly examined next.

\subsection{Comparison with the modified GPT cryptosystem}
\label{sec:GPT}

The original GPT cryptosystem has been the object of Gibson's attack.
To counter such an attack, in \cite{Ourivski2003} a variant including a 
column scrambler in place of the permutation matrix has been
proposed.

Apart from the code extension and the inclusion of an additive
distortion matrix, in the modified GPT cryptosystem the public generator matrix is 
obtained through right-multiplication by a non-singular matrix that is not 
necessarily a permutation matrix.
So, in principle, it seems the same idea of using a more general transformation
matrix as in the proposed cryptosystem.
However, in order to preserve the ability to correct the intentional
error vectors, the GPT cryptosystem works in the rank metric domain 
and adopts rank distance codes, like Gabidulin codes.

Unfortunately, the properties of Gabidulin codes make it possible to
exploit the effect of the Frobenius automorphism on the public generator matrix
in order to mount a polynomial-time attack \cite{Overbeck2008}.
Recently, it has been shown that this attack can be avoided \cite{Rashwan2010},
but the cryptosystem still needs to work with rank distance codes.
Differently from the GPT cryptosystem, the proposed solution is able
to exploit Hamming distance codes, that: i) are more widespread than
rank distance codes, ii) can be chosen to have convenient properties or 
structure, like GRS codes, and iii) may take advantage of many efficient codec 
implementations that are already available.

\subsection{Comparison with full-decoding cryptosystems}
\label{sec:FullDecoding}

The main idea behind full-decoding cryptosystems in \cite{Kabatiansky2005} 
is to let the intentional error vectors have any arbitrary weight. This way,
an attacker would be forced to try full-decoding of the public code, that 
is known to be an NP-complete task.
Obviously, the legitimate receiver must be able to decode any intentional error vector with reasonable complexity; so, the 
problem of full decoding must be transformed from a one-way function to a trapdoor function. For this purpose, the main 
idea is to use a transformation that maps a set of error vectors with weight $\leq t$ 
into a set of arbitrary weight intentional error vectors.

If this transformation is represented by the $n \times n$ matrix $\mathbf{M}$, the public code (as proposed first 
in \cite{Kabatiansky2005}) would be $\mathbf{G}' = \mathbf{G} \cdot \mathbf{M}$.
The basic point for obtaining a trapdoor function is to make Alice use only those error vectors that can be expressed 
as $\mathbf{e}' = \mathbf{e} \cdot \mathbf{M}$, where $\mathbf{e}$ is a weight-$t$ error vector. This way, when Bob uses 
the inverse of the secret matrix $\mathbf{M}$ to invert the transformation, he re-maps each arbitrary weight error vector 
into a correctable error vector. Unauthorized users would instead be forced to try full-decoding over arbitrary weight 
error vectors; so, the trapdoor is obtained.

The set of intentional error vectors used in full-decoding cryptosystems is
not the set (or a subset) of the correctable error vectors,
as in the proposed cryptosystem, but a transformed version of it. 
In fact, the purpose of full-decoding cryptosystems is to 
increase the security level with respect to the McEliece cryptosystem by relying on a problem that is harder to solve.
In order to exploit the full-decoding problem, Alice must use for encryption only those error vectors that can be anti-transformed 
into correctable error vectors. 
So, some information on the transformation used to originate them must be disclosed.
A solution is that the first $p < n$ rows of $\mathbf{M}$ are made public \cite{Kabatiansky2005}. 
However, it has been proved that, this way, the security reduces to 
that of the original McEliece cryptosystem, and an attacker does not have to attempt full-decoding, but only normal decoding. 

Further variants aim at better hiding the secret transformation matrix in its disclosed version \cite{Kabatiansky2005}.
In the last variant, a generator matrix of a maximum distance-$t$ anticode is used to hide the secret transformation. 
This way, after inverting the secret transformation, the error vector remains correctable for the legitimate receiver.
To our knowledge, the latter version has never been proved to be insecure nor to reduce to the same problem of the original McEliece cryptosystem. 
However, the construction based on anticodes seems unpractical.

Differently from full-decoding cryptosystems,
our proposal still relies on the same problem as the original McEliece cryptosystem 
(that is, normal decoding); so, no transformation is performed over the correctable
random error vectors, but we need, at most, only a selection of them.
For this reason, the information leakage on the secret transformation matrix that 
is needed in the proposed cryptosystem is considerably smaller with respect to what 
happens in full-decoding cryptosystems.

\section{Attacks against the proposed cryptosystem}
\label{sec:Attacks}

A first concern about the proposed cryptosystem is to verify that it is actually able
to provide increased key security, with respect to previous variants of the
McEliece cryptosystem, in such a way as to allow the use of widespread families
of codes (like GRS codes) without
incurring in the attacks that have prevented their use up to now.

From the comparison with the variants described in Sections \ref{sec:GPT} and
\ref{sec:FullDecoding}, we infer that previous attacks targeted to those cryptosystems
do not succeed against the proposed one, due to the differences in the family of codes 
used and in the information leakage on the secret transformation.
Concerning the latter point, we observe that, even if the whole matrix $\mathbf{R}$
(and not only the vector $\mathbf{a}$) were public, an attacker would not gain much 
information. 
In fact, in this case, he could compute $\mathbf{x} \cdot \mathbf{R} = \mathbf{u} \cdot \mathbf{G}' \cdot \mathbf{R}$.
However, for the choices of the parameters we consider, $\mathbf{R}$ has rank $\ll n$, so $\mathbf{G}' \cdot \mathbf{R}$
is not invertible and recovering $\mathbf{u}$ is not possible.

The most general attack procedures against code-based cryptosystems, hence against our proposed solution,
are those techniques that attempt 
information set decoding (ISD) on the public code; so we estimate the security level 
of the proposed cryptosystem against this kind of attacks.
Actually, there is no guarantee that the public code, defined through
the generator matrix \eqref{eq:Gprime} or, equivalently, the parity-check
matrix \eqref{eq:NiederreiterKey}, maintains the same minimum distance and
error correction capability of the secret code.
Since the private code has very good distance properties, and the transformation
matrix is randomly chosen, the public code will most probably have worse minimum
distance than the private one.
So, in estimating the security level as the work factor (WF) of ISD attacks, 
we make the pessimistic assumption that the public code is still able to
correct all intentional errors.

More specific attack techniques are those aimed at exploiting the particular 
structure of the adopted codes. In this case, a necessary condition to perform the attack
is the ability to distinguish the public code matrix from a random matrix.
If the attacker cannot distinguish the complete random case from the implemented one, he is
forced to use ISD attack procedures in place of specific ones.
Various distinguisher techniques are used against specific codes: a notable one is that presented in \cite{Faugere2010};
in this case the authors propose a polynomial algorithm to distinguish high rate alternant codes (Goppa codes are
alternant codes) from random codes. Since the existence of a distinguisher attack would be more effective than ISD attacks, we discuss this issue first, in the next subsection.

\subsection{Distinguisher attacks}
\label{subsec:Distinguisher}

We will analyze two possible kinds of distinguisher for the case of GRS codes adopted as secret codes.
The first one is that proposed in \cite{Faugere2010}, that is able
to distinguish matrices in the classic Goppa code-based McEliece cryptosystem and CFS signature scheme \cite{Courtois2001}, for certain 
system parameters. The second one derives from \cite{Gauthier2012, Couvreur2013}, where the authors focus just on the GRS codes.
Both of them do not succeed in breaking the system we propose in its general version, but, as often happens when dealing with 
distinguisher attacks, the second one forces a particular choice of the system parameters, in the same manner as the first one 
forces certain parameters for Goppa codes.
We notice that a distinguisher, able to discriminate between a random matrix and the generator (or parity check) matrix of the public code, gives
a clue regarding some possible vulnerabilities but does not define an attack procedure, in strict sense.
However, in the particular case of GRS based matrices, it is possible to derive an attack on the basis of a modified distinguisher \cite{Gauthier2012, Couvreur2013}.
Before introducing the attack, we remark that, since the dual space of a GRS code is still a GRS code, the parity check matrix of a GRS code is still a generator matrix of a GRS code having dimension and redundancy inverted with respect to the first one. This implies that the following procedure can be applied both to the McEliece and Niederreiter versions of the system. For the sake of clarity, we will refer to $\mathbf{G}$ as a generic GRS generator matrix, also in accordance with the notation used in \cite{Gauthier2012}.

We define a \emph{Distinguisher Attack Procedure} (DAP) through:
\begin{itemize}
\item the public code $\mathcal{C}_{pub}$ described by the public key matrix $\mathbf{G'}$
\item the code $\mathcal{C}$ whose generator matrix is $\mathbf{G \cdot T^{-1}}$
\item the matrix $\mathbf{R \cdot T^{-1}}=\mathbf{B'^{T} \cdot A'}$, having rank equal to $z$
\item the matrix $\mathbf{P=I+R \cdot T^{-1}}$, where $\mathbf{I}$ is the identity matrix
\item the matrix $\mathbf{\Lambda}=\mathbf{P^{-1} \cdot B'^{T}}$
\item the code $\mathcal{C}_{\Lambda^{\bot}}=\mathcal{C}\cap<\mathbf{\Lambda}>^{\bot}$, where $<\mathbf{\Lambda}>^{\bot}$ is the space having $\mathbf{\Lambda}$ as parity check matrix.
\end{itemize} 
$\mathbf{A'}$ and $\mathbf{B'}$ are $n\times z$ matrices, whose existence is ensured by setting $\mathbf{Q}=\mathbf{R+T}$.
It is possible to show that $\mathcal{C}_{\Lambda^{\bot}}$ is a large subspace of both $\mathcal{C}_{pub}$ and $\mathcal{C}$.
So, in the case of $m=1$, $\mathcal{C}_{\Lambda^{\bot}}$ is a large subspace of a permuted version of the secret code. Knowing this 
subspace could allow the attacker to use the algorithm introduced in \cite{Wieschebrink2010} for recovering the secret code, once he has 
recovered the description of $\mathcal{C}_{\Lambda^{\bot}}$ as GRS code, using the algorithm presented in \cite{Sidelnikov1992}.
If the attacker is able to distinguish between the vectors belonging to $\mathcal{C}_{pub}$, but not to $\mathcal{C}_{\Lambda^{\bot}}$, and 
those belonging to $\mathcal{C}_{pub}$ and to $\mathcal{C}_{\Lambda^{\bot}}$ he could hence recover the secret key.

\subsubsection{Alternant distinguisher}

The idea behind \cite{Faugere2010} is to consider the dimension of the solution space of a linear system
deduced from the polynomial system describing the alternant (Goppa) code by a linearization technique which introduces many unknowns.
The solution of this linearized system is indeed an algebraic attack against particular instances of the McEliece cryptosystem
(those having very structured matrices like quasi-cyclic or quasi-dyadic codes, that allow to reduce the complexity of the
linearized system). 
However this attack is not feasible in the general case, that is the case of classic Goppa codes with no further structure.

The authors propose not to solve the system, but rather to consider the dimension of its solution space in such a way as to distinguish systems
induced from an alternant code, a Goppa code or a random one.
However, this distinguisher is ineffective against the system we propose, since:

\begin{itemize}
\item it is not able to distinguish the public key matrix of the proposed cryptosystem from a randomly generated one, that is, our keys are resistant to this distinguisher since they are not generator matrices of alternant or GRS codes (this is due to the fact that $\mathbf{Q}$ is not a permutation matrix);
\item it does not allow to mount a DAP: the distinguisher cannot work on subspaces of the code, so it is unable to recover the subspace the attacker needs.
\end{itemize}

\subsubsection{GRS code-based distinguisher}
\label{Distinguisher}

Let us denote by $\star$ the so-called \emph{star product} \cite{Marquez2012}. 
Given $\mathbf{a}=[a_1, a_2, \dots, a_n]$ and $\mathbf{b}=[b_1, b_2, \dots, b_n]$, we have $\mathbf{a}\star\mathbf{b}=[a_1b_1, a_2b_2, \dots, a_nb_n]$.
Using this star-product on the elements of the public code, another distinguisher was devised in \cite{Gauthier2012, Couvreur2013}, specifically designed to obtain a subcode needed to attack the system through a DAP.

The key idea is to choose $3$ random codewords $\mathbf{c_1,c_2,c_3}$ of the code described by the public generator matrix and compute all
the possible star products $\mathbf{g_i}\star \mathbf{c_j}$, where $\mathbf{g_i}$ is one of the rows in $\mathbf{G'}$, with $1\le i \le k$ and $1\le j \le 3$.

It is possible to verify that, if $\left\{\mathbf{c_1,c_2,c_3}\right\}\in \mathbf{C}_{\Lambda^{\bot}}$, the dimension of the space described by
$\mathbf{G'}\star \left\{\mathbf{c_1,c_2,c_3}\right\}$ is almost always equal to (or very close to) $2k+2$; otherwise, when at least one
$\mathbf{c_j}\notin \mathbf{C}_{\Lambda^{\bot}}$, the dimension is equal to (or very close to) $3k-3$. For the sake of simplicity, we call 
$\mathcal{D}_{\mathbf{\Lambda}}$ and $\mathcal{D}_{rand}$ the dimension of the distinguisher space in the two cases.
We note that the DAP in \cite{Gauthier2012, Couvreur2013} can be applied to a code having rate $<0.5$ or to its dual if the rate is $>0.5$; so, $k$
has to be replaced by $r=n-k$ for the cases we consider, where the rates are greater than $0.5$.

Actually, by assuming $z = 1$ and $m = 1$ the complexity of the distinguisher phase of the attack, hence not considering the subsequent subcode recovering phase and the Sidelnikov-Shestakov attack, 
is $O(nk^2q^3)$, where $q$ is the cardinality of the field; so, it seems feasible for any reasonable choice of the parameters.
 
Based on this fact, in \cite{Gauthier2012} it has been demonstrated that a DAP is feasible when $z = 1$ and $m = 1$. This is obviously a very particular choice. In the following, instead, we will consider more general cases with $z > 1$ and $m > 1$: one or both of these choices make the system immune to this kind of attacks.
In fact, two possible countermeasures to this DAP can be devised, both based on an increase of $z$ and/or $m$; the first one imposes to increase the decoding complexity,
while the second one comes for free, but requires good error correction capabilities.

The probability to find, in a single attempt, a set of three vectors belonging to $\mathbf{C}_{\Lambda^{\bot}}$
is $\frac{1}{q^{3z}}$. This means that increasing $z$ yields a large increase in the distinguisher phase work factor.
Another strategy, that allows to avoid the DAP regardless of its work factor, is to increase the value of $m$.
In fact, we have verified numerically that increasing from $1$ to $2$ the weight of a single row of matrix $\mathbf{T}$ has the effect of increasing also
$\mathcal{D}_{\mathbf{\Lambda}}$ by the same quantity.
We have verified that this effect remains even when the weight-$2$ rows of $\mathbf{T}$ have one of the two non-zero symbols concentrated in a
small number of columns, which is a desirable feature in the design of $\mathbf{T}$, as explained in Section \ref{subsec:T}.

When $\mathcal{D}_{\mathbf{\Lambda}} = \mathcal{D}_{rand}$, the distinguisher fails, since there is no dimension difference between the space the attacker needs to mount the DAP and the public key space.
The condition $\mathcal{D}_{\mathbf{\Lambda}} = \mathcal{D}_{rand}$ can be achieved by adding $3r-3-(2r+2)=r-5$ non-null elements to the matrix $\mathbf{T}$. In turn, this can be accomplished by setting $m\ge1+\frac{r-5}{n}$.   
On the other hand, in \cite{Gauthier2012} the authors notice a non-negligible probability that
$\mathcal{D}_{\mathbf{\Lambda}}$ is slightly smaller than its expected value; so, it can be useful to increase the value of $m$ such that $m\ge1+\frac{r-3}{n}$.
Actually, this is only a precautionary condition, since, in all our tests, the defect in $\mathcal{D}_{\mathbf{\Lambda}}$ or in $\mathcal{D}_{rand}$ was never noticed.
 
Let $t=t_{GRS}$ be the correction capability of the secret GRS code, and $t_{pub}=\lfloor \frac{t}{m} \rfloor$ the number of intentional errors in the encrypted message (for the McEliece version) or the number of errors generating the transmitted syndrome (in the Niederreiter version).

The choice of $m>1$, which is needed to avoid attacks based on distinguishers, affects the WF of the ISD attack techniques that will be presented in Section \ref{subsec:ISDnonbinary}, since, for a fixed error correction capability of the private code, the number of intentional errors to be added during encryption decreases as $m$ increases over $1$.

\subsection{ISD attacks}
\label{subsec:ISDnonbinary}

The ISD attack is non-polynomial in the code dimension, since it aims at decoding a random linear 
code without exploiting any structural property (even if present) and this task is notoriously non-polynomial.

The complexity of ISD algorithms depends on the actual number of errors added to a codeword (besides the cardinality of the field, the code length and dimension),
and not on the code correction capability; so, it is crucial to assess the number of errors the algorithm is searching for.
For such reason, we investigate whether the constraints that may be imposed on the intentional error vectors
in the proposed cryptosystem have any consequences on its security.
For this purpose, the approach we adopt consists in considering a reduced 
number of intentional errors in the WF computations, that is, $t_{pub}' = t_{pub} - z$.
This approach is conservative in the sense that we assume that the attacker exactly knows both the position 
and the value of $z$ errors, while he actually knows only their values. 

In \cite{Bernstein2008} the authors propose some smart speedup techniques to 
reduce the work factor of Stern's algorithm for ISD over the binary field, this way obtaining 
a theoretical WF close to $2^{60}$ for the original set of parameters ($n=1024$, $k=524$, $t=50$).
As a consequence, the authors consider some new set of system parameters in order
to increase the security level.
One of the biggest improvements presented in \cite{Bernstein2008} is a smart way to find 
$k$ independent columns in the public generator matrix at each iteration without performing Gaussian reduction 
on all such columns. A further improvement consists in the pre-computation of the sum of some rows during the reduction.

In \cite{Peters2010}, the algorithm is generalized to work over larger fields, and
it is shown that the speedups introduced in \cite{Bernstein2008} are mostly efficient on very small fields.
As it can be argued from the table available in \cite{Peters2010website}, for $q > 16$ 
the maximum values of the speedup parameters are $c=2$, $s=1$,
where $c$ represents the number of columns to be changed in the case an iteration fails and $s$ is the number of 
rows in a single pre-sum ($1$ means no speedup).
So, for large fields, 
these speedups are not relevant and the algorithm is quite similar to Stern's one. The difference 
relies on guessing not only $p$ error positions but also $p$ error values in the $k$ independent columns, due to the field 
cardinality.

Concerning ISD over the binary field, several advances have recently appeared in the literature \cite{May2011, Bernstein2011a, Becker2012},
which are able to reduce the attack work factor.
Non-asymptotic estimates of the work factor of the most recent algorithm \cite{Becker2012} are reported in \cite{Misoczki2012, Hamdaoui2013}.
Unfortunately, there is no straightforward generalization of this algorithm to work over non-binary fields.
Therefore, we adopt a heuristic and conservative approach to take into account the possible
improvement coming from its generalization to non-binary fields.
It relies on the following observations:
\begin{itemize}
\item For a fixed set of parameters (code length and rate, and number of errors to correct), the work factor of the algorithm in \cite{Peters2010}
is moderately affected by the field size. For example, for codes with $(n,k)$ as in Table \ref{tab:ISDGF547}, $z=1$, and a number of errors equal to $t_{pub}-1$,
passing from $\mathbb{F}_{547}$ to $\mathbb{F}_{2}$ gives a maximum reduction of the work factor in the order of $2^{10}$.
The same holds for codes with the parameters in Table \ref{tab:ISDGF347}, passing from $\mathbb{F}_{347}$ to $\mathbb{F}_{2}$.
This conclusion results from Tables \ref{tab:ISDGF547to2} and \ref{tab:ISDGF347to2}, where we report the values of
the ISD work factor computed according to \cite{Peters2010} for these two sets of code parameters, as a function of the number of errors and the field size.
Such values of work factor have been computed through the PARI/GP script available in \cite{Peters2010website}.
\item By considering the most recent ISD variant \cite{Becker2012} and estimating its work factor as in \cite{Misoczki2012},
we obtain that, for the binary case, a work factor reduction in the order of $2^9$ or less results with respect to the approach in \cite{Peters2010},
when codes with the parameters in Tables \ref{tab:ISDGF547} and \ref{tab:ISDGF347} are considered.
This also results from Tables \ref{tab:ISDGF547to2} and \ref{tab:ISDGF347to2}, where we report the values of the ISD work factor computed 
according to \cite{Becker2012, Misoczki2012}, for the binary case, as a function of the number of errors.
\end{itemize}
Based on these considerations, we assume that, if a generalization of the algorithm in \cite{Becker2012} to non-binary fields were found,
it would result in a work factor reduction in the order of $2^9$ or less with respect to the algorithm in \cite{Peters2010}, for the parameters we consider.

\begin{table}[ht]
\renewcommand{\arraystretch}{1.3}
\caption{Work factor ($\mathrm{log}_2$) of ISD attacks on GRS codes with $n=546$, defined over several finite fields, for $m=1+\frac{r-3}{n}$ and $z=1$.}
\label{tab:ISDGF547to2}
\centering
\scriptsize\begin{tabular}{|c||@{\hspace{0.5mm}}c@{\hspace{0.5mm}}|@{\hspace{0.5mm}}c@{\hspace{0.5mm}}|@{\hspace{0.5mm}}c@{\hspace{0.5mm}}|@{\hspace{0.5mm}}c@{\hspace{0.5mm}}|@{\hspace{0.5mm}}c@{\hspace{0.5mm}}|@{\hspace{0.5mm}}c@{\hspace{0.5mm}}|@{\hspace{0.5mm}}c@{\hspace{0.5mm}}|@{\hspace{0.5mm}}c@{\hspace{0.5mm}}|@{\hspace{0.5mm}}c@{\hspace{0.5mm}}|@{\hspace{0.5mm}}c@{\hspace{0.5mm}}|@{\hspace{0.5mm}}c@{\hspace{0.5mm}}|@{\hspace{0.5mm}}c@{\hspace{0.5mm}}|@{\hspace{0.5mm}}c@{\hspace{0.5mm}}|@{\hspace{0.5mm}}c@{\hspace{0.5mm}}|}
\hline
$k$ & $428$ & $420$ & $412$ & $404$ & $396$ & $388$ & $380$ & $372$ & $364$ & $356$ & $348$ & $340$ & $332$ & $324$ \\
\hline
$t_{pub}$ & $48$ & $51$ & $54$ & $56$ & $59$ & $61$ & $63$ & $66$ & $68$ & $70$ & $72$ & $75$ & $77$ & $79$ \\
\hline
$\mathrm{WF}_{z=1}(\mathbb{F}_{547})$ \cite{Peters2010} & $131.1$ & $133.8$ & $136.1$ & $135.7$ & $137.6$ & $136.9$ & $136.2$ & $137.5$ & $136.6$ & $135.5$ & $134.4$ & $135.1$ & $133.8$ & $132.4$ \\
\hline
$\mathrm{WF}_{z=1}(\mathbb{F}_{256})$ \cite{Peters2010} & $130.4$ & $132.9$ & $135.1$ & $134.6$ & $136.4$ & $135.7$ & $134.9$ & $136.1$ & $135.1$ & $134.0$ & $132.9$ & $133.4$ & $132.1$ & $130.6$ \\
\hline
$\mathrm{WF}_{z=1}(\mathbb{F}_{128})$ \cite{Peters2010} & $128.5$ & $131.1$ & $133.3$ & $132.8$ & $134.6$ & $133.9$ & $133.1$ & $134.3$ & $133.3$ & $132.3$ & $131.1$ & $131.6$ & $130.3$ & $128.8$ \\
\hline
$\mathrm{WF}_{z=1}(\mathbb{F}_{64})$ \cite{Peters2010} & $126.9$ & $129.4$ & $131.6$ & $131.1$ & $132.9$ & $132.2$ & $131.4$ & $132.6$ & $131.6$ & $130.5$ & $129.4$ & $129.9$ & $128.6$ & $127.1$ \\
\hline
$\mathrm{WF}_{z=1}(\mathbb{F}_{32})$ \cite{Peters2010} & $125.2$ & $127.7$ & $130.0$ & $129.5$ & $131.3$ & $130.6$ & $129.8$ & $131.0$ & $130.0$ & $129.0$ & $127.8$ & $128.4$ & $127.0$ & $125.6$ \\
\hline
$\mathrm{WF}_{z=1}(\mathbb{F}_{16})$ \cite{Peters2010} & $123.7$ & $126.3$ & $128.5$ & $128.1$ & $129.9$ & $129.2$ & $128.4$ & $129.7$ & $128.7$ & $127.7$ & $126.5$ & $127.1$ & $125.8$ & $124.4$ \\
\hline
$\mathrm{WF}_{z=1}(\mathbb{F}_{2})$ \cite{Peters2010} & $123.4$ & $125.9$ & $128.1$ & $127.4$ & $129.2$ & $128.3$ & $127.3$ & $128.6$ & $127.4$ & $126.2$ & $124.9$ & $125.5$ & $124.0$ & $122.5$ \\
\hline
$\mathrm{WF}_{z=1}(\mathbb{F}_{2})$ \cite{Becker2012, Misoczki2012} & $115.2$ & $117.6$ & $119.8$ & $119.3$ & $121.0$ & $120.1$ & $119.1$ & $120.2$ & $119.0$ & $117.8$ & $116.4$ & $116.9$ & $115.4$ & $114.0$ \\
\hline
\end{tabular}
\end{table}

\begin{table}[ht]
\renewcommand{\arraystretch}{1.3}
\caption{Work factor ($\mathrm{log}_2$) of ISD attacks on GRS codes with $n=346$, defined over several finite fields, for $m=1+\frac{r-3}{n}$ and $z=1$.}
\label{tab:ISDGF347to2}
\centering
\scriptsize\begin{tabular}{|c||@{\hspace{0.5mm}}c@{\hspace{0.5mm}}|@{\hspace{0.5mm}}c@{\hspace{0.5mm}}|@{\hspace{0.5mm}}c@{\hspace{0.5mm}}|@{\hspace{0.5mm}}c@{\hspace{0.5mm}}|@{\hspace{0.5mm}}c@{\hspace{0.5mm}}|@{\hspace{0.5mm}}c@{\hspace{0.5mm}}|@{\hspace{0.5mm}}c@{\hspace{0.5mm}}|@{\hspace{0.5mm}}c@{\hspace{0.5mm}}|@{\hspace{0.5mm}}c@{\hspace{0.5mm}}|@{\hspace{0.5mm}}c@{\hspace{0.5mm}}|@{\hspace{0.5mm}}c@{\hspace{0.5mm}}|@{\hspace{0.5mm}}c@{\hspace{0.5mm}}|@{\hspace{0.5mm}}c@{\hspace{0.5mm}}|@{\hspace{0.5mm}}c@{\hspace{0.5mm}}|@{\hspace{0.5mm}}c@{\hspace{0.5mm}}|}
\hline
$k$ & $284$ & $276$ & $268$ & $260$ & $252$ & $244$ & $236$ & $228$ & $220$ & $212$ & $204$ & $196$ & $188$ & $180$ & $172$ \\
\hline
$t_{pub}$ & $26$ & $29$ & $32$ & $34$ & $37$ & $39$ & $42$ & $44$ & $46$ & $48$ & $50$ & $52$ & $54$ & $56$ & $58$ \\
\hline
$\mathrm{WF}_{z=1}(\mathbb{F}_{347})$ \cite{Peters2010} & $82.4$ & $85.9$ & $88.8$ & $88.8$ & $90.9$ & $90.5$ & $92.0$ & $91.1$ & $90.2$ & $89.1$ & $87.9$ & $86.6$ & $85.1$ & $83.6$ & $82.0$ \\
\hline
$\mathrm{WF}_{z=1}(\mathbb{F}_{256})$ \cite{Peters2010} & $82.3$ & $85.8$ & $88.7$ & $88.7$ & $90.8$ & $90.4$ & $91.9$ & $91.0$ & $90.1$ & $89.0$ & $87.8$ & $86.5$ & $85.1$ & $83.5$ & $81.9$ \\
\hline
$\mathrm{WF}_{z=1}(\mathbb{F}_{128})$ \cite{Peters2010} & $81.9$ & $85.2$ & $87.9$ & $87.8$ & $89.8$ & $89.2$ & $90.6$ & $89.6$ & $88.6$ & $87.4$ & $86.2$ & $84.8$ & $83.3$ & $81.6$ & $79.9$ \\
\hline
$\mathrm{WF}_{z=1}(\mathbb{F}_{64})$ \cite{Peters2010} & $80.3$ & $83.6$ & $86.3$ & $86.2$ & $88.2$ & $87.7$ & $89.0$ & $88.1$ & $87.0$ & $85.9$ & $84.6$ & $83.2$ & $81.7$ & $80.0$ & $78.3$ \\
\hline
$\mathrm{WF}_{z=1}(\mathbb{F}_{32})$ \cite{Peters2010} & $78.4$ & $81.6$ & $84.4$ & $84.3$ & $86.3$ & $85.7$ & $87.1$ & $86.2$ & $85.1$ & $84.0$ & $82.7$ & $81.3$ & $79.9$ & $78.3$ & $76.6$ \\
\hline
$\mathrm{WF}_{z=1}(\mathbb{F}_{16})$ \cite{Peters2010} & $76.9$ & $80.2$ & $83.0$ & $82.8$ & $84.9$ & $84.3$ & $85.7$ & $84.8$ & $83.8$ & $82.7$ & $81.5$ & $80.2$ & $78.8$ & $77.3$ & $75.7$ \\
\hline
$\mathrm{WF}_{z=1}(\mathbb{F}_{2})$ \cite{Peters2010} & $74.9$ & $75.2$ & $81.6$ & $81.2$ & $83.2$ & $82.4$ & $83.8$ & $82.8$ & $81.6$ & $80.3$ & $79.0$ & $77.6$ & $76.0$ & $74.4$ & $72.6$ \\
\hline
$\mathrm{WF}_{z=1}(\mathbb{F}_{2})$ \cite{Becker2012, Misoczki2012} & $68.2$ & $71.6$ & $74.2$ & $73.7$ & $75.6$ & $74.8$ & $76.0$ & $74.9$ & $73.7$ & $72.4$ & $70.9$ & $69.4$ & $68.2$ & $66.2$ & $64.5$ \\
\hline
\end{tabular}
\end{table}

\subsection{Numerical examples}
\label{subsec:Examples}

In Tables \ref{tab:ISDGF547} and \ref{tab:ISDGF347} we report some values of the ISD attack WF, when using GRS codes in the
variant of the McEliece cryptosystem we propose, with $m=1+\frac{r-3}{n}$ and $z=1,2,3,4$.
They were computed through the PARI/GP script available in \cite{Peters2010website}, that allows the estimation of the security level
based on the algorithm in \cite{Peters2010}.
The reported WFs are the lowest ones obtained for each set of parameters.

\begin{table}[ht]
\renewcommand{\arraystretch}{1.3}
\caption{Work factor ($\mathrm{log}_2$) of ISD attacks estimated as in \cite{Peters2010} for GRS codes with $n=546$, defined over $\mathbb{F}_{547}$, and $m=1+\frac{r-3}{n}$, $z=1,2,3,4$.}
\label{tab:ISDGF547}
\centering
\scriptsize\begin{tabular}{|c||@{\hspace{0.5mm}}c@{\hspace{0.5mm}}|@{\hspace{0.5mm}}c@{\hspace{0.5mm}}|@{\hspace{0.5mm}}c@{\hspace{0.5mm}}|@{\hspace{0.5mm}}c@{\hspace{0.5mm}}|@{\hspace{0.5mm}}c@{\hspace{0.5mm}}|@{\hspace{0.5mm}}c@{\hspace{0.5mm}}|@{\hspace{0.5mm}}c@{\hspace{0.5mm}}|@{\hspace{0.5mm}}c@{\hspace{0.5mm}}|@{\hspace{0.5mm}}c@{\hspace{0.5mm}}|@{\hspace{0.5mm}}c@{\hspace{0.5mm}}|@{\hspace{0.5mm}}c@{\hspace{0.5mm}}|@{\hspace{0.5mm}}c@{\hspace{0.5mm}}|@{\hspace{0.5mm}}c@{\hspace{0.5mm}}|@{\hspace{0.5mm}}c@{\hspace{0.5mm}}|}
\hline
$k$ & $428$ & $420$ & $412$ & $404$ & $396$ & $388$ & $380$ & $372$ & $364$ & $356$ & $348$ & $340$ & $332$ & $324$ \\
\hline
$t_{GRS}$ & $59$ & $63$ & $67$ & $71$ & $75$ & $79$ & $83$ & $87$ & $91$ & $95$ & $99$ & $103$ & $107$ & $111$ \\
\hline
$m$ & $1.211$ & $1.225$ & $1.240$ & $1.255$ & $1.269$ & $1.284$ & $1.299$ & $1.313$ & $1.328$ & $1.342$ & $1.357$ & $1.372$ & $1.386$ & $1.401$ \\
\hline
$t_{pub}$ & $48$ & $51$ & $54$ & $56$ & $59$ & $61$ & $63$ & $66$ & $68$ & $70$ & $72$ & $75$ & $77$ & $79$ \\
\hline
$\mathrm{WF}_{z=1}$ & $131.1$ & $133.8$ & $136.1$ & $135.7$ & $137.6$ & $136.9$ & $136.2$ & $137.5$ & $136.6$ & $135.5$ & $134.4$ & $135.1$ & $133.8$ & $132.4$ \\
\hline
$\mathrm{WF}_{z=2}$ & $128.4$ & $131.1$ & $133.5$ & $133.3$ & $135.2$ & $134.7$ & $134.0$ & $135.4$ & $134.5$ & $133.6$ & $132.6$ & $133.3$ & $132.0$ & $130.7$ \\
\hline
$\mathrm{WF}_{z=3}$ & $125.7$ & $128.5$ & $131.0$ & $130.8$ & $132.9$ & $132.4$ & $131.9$ & $133.3$ & $132.5$ & $131.7$ & $130.7$ & $131.5$ & $130.3$ & $129.1$ \\
\hline
$\mathrm{WF}_{z=4}$ & $123.0$ & $125.9$ & $128.5$ & $128.4$ & $130.6$ & $130.2$ & $129.7$ & $131.3$ & $130.6$ & $129.8$ & $128.9$ & $129.7$ & $128.6$ & $127.4$ \\
\hline
\end{tabular}
\end{table}

\begin{table}[ht]
\renewcommand{\arraystretch}{1.3}
\caption{Work factor ($\mathrm{log}_2$) of ISD attacks estimated as in \cite{Peters2010} for GRS codes with $n=346$, defined over $\mathbb{F}_{347}$, and $m=1+\frac{r-3}{n}$, $z=1,2,3,4$.}
\label{tab:ISDGF347}
\centering
\scriptsize\begin{tabular}{|c||@{\hspace{0.5mm}}c@{\hspace{0.5mm}}|@{\hspace{0.5mm}}c@{\hspace{0.5mm}}|@{\hspace{0.5mm}}c@{\hspace{0.5mm}}|@{\hspace{0.5mm}}c@{\hspace{0.5mm}}|@{\hspace{0.5mm}}c@{\hspace{0.5mm}}|@{\hspace{0.5mm}}c@{\hspace{0.5mm}}|@{\hspace{0.5mm}}c@{\hspace{0.5mm}}|@{\hspace{0.5mm}}c@{\hspace{0.5mm}}|@{\hspace{0.5mm}}c@{\hspace{0.5mm}}|@{\hspace{0.5mm}}c@{\hspace{0.5mm}}|@{\hspace{0.5mm}}c@{\hspace{0.5mm}}|@{\hspace{0.5mm}}c@{\hspace{0.5mm}}|@{\hspace{0.5mm}}c@{\hspace{0.5mm}}|@{\hspace{0.5mm}}c@{\hspace{0.5mm}}|}
\hline
$k$ & $284$ & $276$ & $268$ & $260$ & $252$ & $244$ & $236$ & $228$ & $220$ & $212$ & $204$ & $196$ & $188$ & $180$ \\
\hline
$t_{GRS}$ & $31$ & $35$ & $39$ & $43$ & $47$ & $51$ & $55$ & $59$ & $63$ & $67$ & $71$ & $75$ & $79$ & $83$ \\
\hline
$m$ & $1.171$ & $1.194$ & $1.217$ & $1.240$ & $1.263$ & $1.286$ & $1.309$ & $1.332$ & $1.355$ & $1.379$ & $1.402$ & $1.425$ & $1.448$ & $1.471$ \\
\hline
$t_{pub}$ & $26$ & $29$ & $32$ & $34$ & $37$ & $39$ & $42$ & $44$ & $46$ & $48$ & $50$ & $52$ & $54$ & $56$ \\
\hline
$\mathrm{WF}_{z=1}$ & $82.4$ & $85.9$ & $88.8$ & $88.8$ & $90.9$ & $90.5$ & $92.0$ & $91.1$ & $90.2$ & $89.1$ & $87.9$ & $86.6$ & $85.1$ & $83.6$ \\
\hline
$\mathrm{WF}_{z=2}$ & $79.4$ & $83.1$ & $86.2$ & $86.3$ & $88.6$ & $88.3$ & $89.9$ & $89.2$ & $88.3$ & $87.4$ & $86.3$ & $85.1$ & $83.7$ & $82.3$ \\
\hline
$\mathrm{WF}_{z=3}$ & $76.4$ & $80.3$ & $83.6$ & $83.9$ & $86.3$ & $86.1$ & $87.9$ & $87.3$ & $86.5$ & $85.7$ & $84.7$ & $83.6$ & $82.3$ & $80.9$ \\
\hline
$\mathrm{WF}_{z=4}$ & $73.5$ & $77.6$ & $81.0$ & $81.5$ & $84.0$ & $84.0$ & $85.8$ & $85.4$ & $84.7$ & $84.0$ & $83.1$ & $82.1$ & $80.9$ & $79.6$ \\
\hline
\end{tabular}
\end{table}

Based on Tables \ref{tab:ISDGF547} and \ref{tab:ISDGF347}, we can compare the proposed cryptosystem with some
instances of the McEliece/Niederreiter system based on Goppa codes. Two examples are selected below.

\subsubsection{Example 1}

To reach $\mathrm{WF} \geq 2^{80}$, the ($1632$, $1269$) binary Goppa code is suggested in \cite{Bernstein2008}, 
resulting in a public-key size of $460647$ bits (obtained by storing only $k \cdot r$ bits of $\mathbf{H}$ or $\mathbf{G}$).
With the new variant, we can consider from Table \ref{tab:ISDGF347} the GRS code with $n=346$, $k=252$, $t_{GRS}=47$ over $\mathbb{F}_{347}$,
having an estimated WF of $2^{90.9}$ binary operations with $z=1$.
Hence, its security level remains higher than $2^{80}$ even when considering the improvement estimated in Section \ref{subsec:ISDnonbinary} for possible advances in ISD algorithms over non-binary fields.

Since we choose $m=1+\frac{r-3}{n}$, the distinguisher attack is avoided even when $z=1$, and the weight of the intentional error vector is $t_{pub} = 37$.
This way, by adopting the first implementation (see Section \ref{subsec:FirstImpl}), we obtain a public key size of $199899$ bits,
that is about $57\%$ less than in the revised McEliece/Niederreiter cryptosystem \cite{Bernstein2008}.
If we instead adopt the second implementation (see Section \ref{subsec:SecondImpl}),
we also need to store the $1 \times 346$ vector $\mathbf{a}$, with elements over $\mathbb{F}_{347}$.
This would increase the public key size by $2920$ bits, that is not a significant change.

\subsubsection{Example 2}

To reach $\mathrm{WF} \geq 2^{128}$, the ($2960$, $2288$) binary Goppa code is suggested in \cite{Bernstein2008}, 
resulting in a public-key size of $1537536$ bits.
For the sake of comparison, we consider from Table \ref{tab:ISDGF547} the GRS code with $n=546, k=396$, defined over $\mathbb{F}_{547}$,
which achieves the security level $2^{137.6}$ for $z=1$.
This value remains higher than $2^{128}$ even when considering the improvement estimated in Section \ref{subsec:ISDnonbinary} for possible advances in ISD algorithms over non-binary fields.

By adopting this code in the Niederreiter version of the first implementation (see Section \ref{subsec:FirstImpl}),
and storing the last $k$ columns of $\mathbf{H}''$, defined by \eqref{eq:NiederreiterKeySystForm},
we obtain a public key size of $540267$ bits, that is about $65\%$ less than in 
the revised McEliece/Niederreiter cryptosystem based on binary Goppa codes \cite{Bernstein2008}.
If we compare this solution with the non-binary Goppa codes proposed in \cite{Bernstein2011}, defined over
fields ranging between $\mathbb{F}_{3}$ and $\mathbb{F}_{32}$, we get a public key size reduction ranging
between $24\%$ and $68\%$ (we also note that in \cite{Bernstein2011} no improvement over the approach \cite{Peters2010}
was taken into account).
Also in this case, we choose $m=1+\frac{r-3}{n}$ and, hence, the distinguisher attack is avoided even when $z=1$.

\subsubsection{Impact of variable $z$}

The value of $z$ plays a role in the ISD WF computation, as mentioned in Section \ref{subsec:ISDnonbinary}. So, it is meaningful to analyze the impact of increasing the value of $z$, under different assumptions for $m$. 
Similarly to what done before for $z = 1$, we can estimate the WF of an ISD attack for different values of $z$.
Results for $z=2,3,4$ are reported in Tables \ref{tab:ISDGF547} and \ref{tab:ISDGF347}.

As we can observe from the tables, a WF decrease in the order of $2^3$ or less occurs each time $z$ is increased by $1$.
So, for the considered parameters, the security level undergoes some variation, as expected. It should be noted, however, that such an approach is very conservative.
To increase both $m$ and $z$ is an unfavorable condition from the key size standpoint since, reducing the number of correctable intentional errors, it forces the user to increase the error correction capability,
by increasing the code length or reducing the code rate.

Generalizing the analysis in Section \ref{Distinguisher}, that is valid for $m = 1$, a lower bound on the complexity of the DAP can be estimated in $k^3q^{3z}$ operations and, for a given $k$, this value increases by $q^3$ for any increase of $z$ by $1$. Hence, it is possible to verify that, with $m=1$ and $z \geq 2$, the DAP has WF $\ge 2^{80}$ when $q \geq 401$, while, for smaller $q$, $z \geq 3$ is needed.
More complex analyses could be developed to improve the mentioned lower bound, which, however, are outside the goals of the present paper. 

Moreover, we notice that increasing $z$ also has detrimental effects on complexity, as we will show in Section \ref{sec:Complexity}.
Hence, it is preferable to make DAPs unfeasible by choosing $m > 1$, rather than $z > 1$.

\section{Key size and complexity}
\label{sec:Complexity}

In this section, we compare the key length and complexity of the proposed system with those of the classical Goppa code-based cryptosystem and of the RSA algorithm.
We refer to the Niederreiter version of both the proposed cryptosystem and of the Goppa code-based solution.

As regards the key length, as already observed, the key of the proposed system is a $k \times r$ matrix of elements in $\mathbb{F}_q$. 
The same holds for the Goppa code-based Niederreiter cryptosystem, with the only difference that the matrix entries are binary, while for RSA the key length
can be estimated as twice the block size, that is, $2n$ \cite{Canteaut1996}.

As regards the complexity, we must decide the convention for measuring the number of operations. According to \cite{Chen2008}, we consider the cost $\cal S$ of one addition between elements of $\mathbb{F}_q$ to be equal to $l = \left\lceil \log_2(q) \right\rceil$ binary operations, while the cost $\cal M$ of one multiplication equals that of $2l$ additions, that is, ${\cal M} = 2l^2$ binary operations. Following \cite{Chen2008}, we also consider that an inversion over $\mathbb{F}_q$ has the same cost as a multiplication, that is, ${\cal M}$ binary operations.

The right (or left, respectively) multiplication of an $x \times y$ matrix by a vector having $w$ non-null elements requires to sum $w$ columns (or rows, respectively) of the matrix, which costs as $(w-1)x{\cal S}$ (or $(w-1)y{\cal S}$, respectively) binary operations.
When working over $\mathbb{F}_q$ with $q > 2$, this quantity must be added with the operations needed to multiply each element of the vector by the corresponding matrix column (or row, respectively), that is, further $wx{\cal M}$ (or $wy{\cal M}$, respectively) binary operations.
Actually, if the matrix is random, we can consider that each column (or row, respectively) has, on average, $\frac{x}{q}$ (or $\frac{y}{q}$, respectively) null elements.
Hence, computing the element-wise sum or product requires, on average, $x\frac{q-1}{q}$ (or $y\frac{q-1}{q}$, respectively)
sums or multiplications.
For the sake of simplicity, we neglect the term $\frac{q-1}{q}$, thus obtaining slightly pessimistic evaluations.

In the Niederreiter cryptosystem, encryption consists in computing \eqref{eq:NiederreiterCiphertext}.
If we consider the systematic version of the key \eqref{eq:NiederreiterKeySystForm} and split the vector $\mathbf{e}$ into its left and right parts, 
$\mathbf{e} = [\mathbf{e}_l | \mathbf{e}_r]$, the encryption function becomes $\mathbf{x} = \mathbf{e}^T_l + \mathbf{H''}_r \cdot \mathbf{e}^T_r$.
Considering, as in \cite{Canteaut1996}, that on average $\mathbf{e}_r$ has weight equal to $w = \frac{k}{n}t_{pub}$,
the encryption step requires $[(w-1) r + t_{pub} - w] {\cal S}$ binary operations when working over $\mathbb{F}_2$.
More precisely, $(w-1)r$ sums come from the computation of $\mathbf{H''}_r \cdot \mathbf{e}^T_r$,
and further $t_{pub} - w$ sums come from the addition of $\mathbf{e}^T_l$.
When working over $\mathbb{F}_q$, with $q > 2$, the number of binary operations becomes $[(w-1) r + t_{pub} - w] {\cal S} + w r {\cal M}$.
Here we do not consider the encoding step needed to map the information vector into a constant weight vector
(and, then, to demap it), which gives a negligible contribution to the total complexity.

Concerning the decryption stage, we refer to the standard GRS syndrome decoding algorithm, whose complexity 
can be easily estimated in closed form \cite{Chen2008}.
This provides a worst-case estimation, since fast implementations exist which are able to achieve significant
complexity reductions \cite{Chen2008}.
Additional gains can also be obtained by novel techniques as in \cite{Elia2011}, \cite{Schipani2011}.
The complexity of the main steps of GRS syndrome decoding can be estimated \cite{Chen2008} in:
$i$) $4t(2t + 2) {\cal M} + 2t(2t + 1) {\cal S}$ binary operations for the the key equation solver,
$ii$) $n(t - 1) {\cal M} + nt {\cal S}$ binary operations for the Chien search, and
$iii$) $(2t^2 + t) {\cal M} + t(2t - 1) {\cal S}$ binary operations for Forney's formula.
Here we do not consider the syndrome computation step, since the ciphertext is already
computed as a syndrome in the Niederreiter cryptosystem.

The Niederreiter cryptosystem also needs to compute the product $\mathbf{S \cdot x}$.
Since $\mathbf{x}$ is a random $r \times 1$ vector over $\mathbb{F}_q$, we can consider it has, at most, weight $r$.
So, computing $\mathbf{S \cdot x}$ requires further $(r-1) r {\cal S}$ binary operations when working over $\mathbb{F}_2$ and
$(r-1) r {\cal S} + r^2 {\cal M}$ binary operations when working over $\mathbb{F}_q$, with $q > 2$.

The system we propose replaces the permutation matrix with a denser transformation matrix, hence Bob must
compute $\mathbf{e}_T = \mathbf{T}^T \cdot \mathbf{e}^T$, which requires further $(t-1) n {\cal S}$ binary operations when working over $\mathbb{F}_2$ and
$(t-1) n {\cal S} + t n {\cal M}$ binary operations when working over $\mathbb{F}_q$, with $q > 2$.
Furthermore, the proposed system requires to perform the guessing stage described in Section \ref{sec:Design}.
In fact, Bob needs to guess the value of $z$ elements of $\mathbb{F}_q$.
We want to stress that there is no need to execute all of the standard decoding operations in decoding the guessed vector; in fact, there is a very high probability that, if the guessed value is wrong, the word we are trying to decode is indeed not decodable at all.
In this case, the first step of the decoding algorithm, that is the key equation solving algorithm, ends with an 
error, and it is useless to continue through the decoding process.
So, only the key equation solver has to be attempted multiple times, while the algorithms to find the roots of the 
locator polynomial and the value of each error are to be executed only once.
In addition, according to \eqref{eq:xsecond-hiding} and \eqref{eq:xsecond}, each guessing attempt requires 
to perform at most $r$ multiplications and $r$ sums between elements of $\mathbb{F}_q$ (the vectors $\mathbf{H \cdot b}^T$
and $\mathbf{H \cdot 1}^T$ can be precomputed only once, before decryption).

Based on the considerations above, the overall decryption complexity for the Niederreiter version of the proposed cryptosystem
can be estimated as:
\begin{equation}
\begin{array}{ll}
D_{GRS} = & \left\{ [4t(2t+2)+r]\frac{q^z}{2} + 2t^2 + (2n+1)t + r^2 - n \right\} {\cal M} \\
         & + \left\{ [2t(2t+1)+r]\frac{q^z}{2} + 2t^2 + (2n-1)t + (r-1)r - n \right\} {\cal S},
\end{array}
\label{eq:DGRS}
\end{equation} 
where the term $\frac{q^z}{2}$ is given by the mean number of attempts needed to find the right guessed set of $z$ values.

In \cite{Canteaut1996}, an estimation of the decryption complexity for the Goppa code-based Niederreiter cryptosystem can be found. It results in:
\begin{equation}
D_{Goppa} = n + 4g^2t^2 + 2g^2t + gn(2t+1) + \frac{r^2}{2},
\label{eq:DGoppa}
\end{equation}
where $g = \log_2(n)$.

The complexity values estimated so far are expressed in terms of binary operations needed to encrypt or decrypt one ciphertext.
We are more interested in computing the complexity per information bit, thus we divide them by the number of information
bits per ciphertext, that is, $\log_2\left[\left( \begin{array}{c} n \\ t_{pub} \end{array} \right)(q-1)^{t_{pub}}\right]$.

A comparison among the proposed system, the binary Goppa code-based Niederreiter algorithm and RSA (whose complexity has been also evaluated in \cite{Canteaut1996}) is shown in Table \ref{tab:ComparisonMcEliece}, for the same parameters considered in Example 2 of Section \ref{subsec:Examples}. The complexity values are given per information bit and the key length is expressed in bits.
The ciphertext and cleartext size (which coincide with $n$ and $k$, respectively) are expressed in bits for the binary Goppa code-based Niederreiter and RSA schemes, while they are in $q$-ary symbols for the proposed GRS code-based solution.

\begin{table}[ht]
\caption{Comparison between the binary Goppa code-based Niederreiter cryptosystem, RSA and the proposed GRS code-based cryptosystem for $128$-bit security.}
\label{tab:ComparisonMcEliece}
\centering
\begin{tabular}{c c c c}
\hline
\hline
& Binary Goppa code-based & RSA & GRS code-based \\
& Niederreiter &   & proposed \\
\hline
\hline
$n$ & $2960$ & $3072$ & $546$ \\
\hline
$k$ & $2288$ & $3072$ & $396$ \\
\hline
Key size & $1537536$ & $6144$ & $540267$\\
\hline
Enc. complexity & $72$ & $5406$ & $1679$ \\
\hline
Dec. complexity & $15302$ & $6643013$ & $3228153$ \\
\hline
\end{tabular}
\end{table}

These results point out that the proposed cryptosystem can be seen as a tradeoff between the classical
binary Goppa code-based Niederreiter cryptosystem and RSA.
In fact, it is able to reduce the key size, by about three times, with respect to the binary Goppa code-based
solution.
This comes at some cost in complexity, which, however, remains lower than for the widely used RSA.
More in detail, the encoding and decoding complexities of the proposed cryptosystem are, respectively,
more than three and two times smaller than for RSA.

\section{Conclusion}
\label{sec:Conclusion}
We have introduced a variant of the McEliece cryptosystem that, by replacing the secret permutation matrix with a more
general transformation matrix, is able to avoid that the public code is permutation-equivalent to the
secret code. This allows to prevent attacks against classical families of codes, like GRS codes, and to
reconsider them as possible good candidates in this framework.

We have proposed some practical implementations of the new cryptosystem, by considering both its
McEliece and Niederreiter variants, and we have addressed some important issues that may influence
their design.

We have also assessed the security level of the proposed cryptosystem, by considering up-to-date attack procedures,
and we have compared it with the classical McEliece cryptosystem and the Niederreiter variant.
Our results show that the proposed solution, by exploiting GRS codes, is able to guarantee an increased security 
level and, at the same time, a considerable reduction in the public key size.
Moreover, for a given security level, the proposed solution exhibits lower complexity than RSA.

\section*{Acknowledgment}

The authors would like to thank Jean-Pierre Tillich and Ayoub Otmani for having
pointed out the subcode vulnerability for the private code.

\end{document}